\documentclass[conference]{IEEEtran}
\IEEEoverridecommandlockouts

\ifCLASSOPTIONcompsoc
  \usepackage[nocompress]{cite}
\else
  \usepackage{cite}
\fi

\ifCLASSINFOpdf
\else
\fi

\usepackage{graphicx}
\usepackage{booktabs} 
\usepackage{adjustbox}
\usepackage{subcaption}
\usepackage{multirow}

\usepackage{amsmath}
\usepackage{amssymb}
\usepackage{xcolor}
\usepackage{breqn,xspace}
\usepackage{bm}
\usepackage{lipsum,multicol}
\usepackage{makecell}

\usepackage{cleveref}

\usepackage{siunitx}

\usepackage{algorithm}
\usepackage{algorithmic}
\usepackage{amsmath}

\newcommand{\name}{SatFed\xspace}

\usepackage{diagbox}
\allowdisplaybreaks[2]
\usepackage{cleveref}
\usepackage{bbm}

\ifodd 1
\definecolor{darkgreen}{RGB}{0,200,0}
\else

\fi

\begin{document}
%
\title{\name: A Resource-Efficient LEO Satellite-Assisted Heterogeneous Federated Learning Framework}
%
%
%
%


\author{Yuxin~Zhang, Zheng~Lin, Zhe~Chen,~\IEEEmembership{Member,~IEEE}, Zihan~Fang, Wenjun~Zhu,
Xianhao Chen,~\IEEEmembership{Member,~IEEE}, \\Jin~Zhao,~\IEEEmembership{Senior Member,~IEEE,} and Yue Gao,~\IEEEmembership{Fellow,~IEEE}
\thanks{Y. Zhang, Z. Lin, Z. Chen, Z. Fang, W. Zhu, J. Zhao and Y. Gao are with the Institute of Space Internet, Fudan University, Shanghai 200438, China (e-mail: yxzhang24@m.fudan.edu.cn; zlin20@fudan.edu.cn; zhechen@fudan.edu.cn; zhfang19@fudan.edu.cn; wenjun@fudan.edu.cn; jzhao@fudan.edu.cn; gao.yue@fudan.edu.cn). Z. Lin is also with the Department of Electrical and Electronic Engineering, University of Hong Kong, Pok Fu Lam, Hong Kong, China.}
\thanks{X. Chen is with the Department of Electrical and Electronic Engineering,
University of Hong Kong, Pok Fu Lam, Hong Kong, China (e-mail: xchen@eee.hku.hk).}
\thanks{\textit{(Corresponding author: Yue Gao)}}
}

\markboth{Journal of \LaTeX\ Class Files,~Vol.~14, No.~8, August~2015}%
{Shell \MakeLowercase{\textit{et al.}}: Bare Advanced Demo of IEEEtran.cls for IEEE Computer Society Journals}
%



\IEEEtitleabstractindextext{%
\begin{abstract}

%
Traditional federated learning (FL) frameworks rely heavily on terrestrial networks, where coverage limitations and increasing bandwidth congestion significantly hinder model convergence. Fortunately,
the advancement of low-Earth orbit (LEO) satellite networks offers promising new communication avenues to augment traditional terrestrial FL. {Despite this potential, the limited satellite-ground communication bandwidth and the heterogeneous operating environments of ground devices—including variations in data, bandwidth, and computing power—pose substantial challenges for effective and robust satellite-assisted FL.}
To address these challenges, we propose \textit{SatFed}, a resource-efficient satellite-assisted heterogeneous FL framework. SatFed implements freshness-based model prioritization queues to optimize the use of highly constrained satellite-ground bandwidth, ensuring the transmission of the most critical models. Additionally, a multigraph is constructed to capture real-time heterogeneous relationships between devices, including data distribution, terrestrial bandwidth, and computing capability. This multigraph enables SatFed to aggregate satellite-transmitted models into peer guidance, enhancing local training in heterogeneous environments. Extensive experiments with real-world LEO satellite networks demonstrate that SatFed achieves superior performance and robustness compared to state-of-the-art benchmarks.

\end{abstract}

}

\maketitle

\IEEEdisplaynontitleabstractindextext

%
\IEEEpeerreviewmaketitle


\section{Introduction}
\label{sec:introduction}
As edge devices like drones, smart cameras, and IoT sensors proliferate~\cite{wu2024s,10175391,wu2022echohand}, they generate vast amounts of data at the network edge~\cite{lin2024efficient,liu2024sesame}. This data drives significant advancements in key domains such as resource management~\cite{ren2022efficient,lin2021spatial}, authentication\cite{wucong2024tifs}, image compression~\cite{huang2024d}, and autonomous driving~\cite{fang2024pacp,hu2023towards,lin2022channel,hu2024collaborative} through machine learning (ML). However, privacy considerations~\cite{dataprivacy,li2024privacy} and bandwidth constraints~\cite{rodio2023federated} render transferring large volumes of data to a cloud server impractical, necessitating the adoption of distributed ML training methods.

To address these challenges, federated learning (FL)~\cite{FedAvg} has emerged as a promising distributed machine learning framework, allowing edge devices to collaboratively train models without sharing local data~\cite{zhang2024fedac,lin2023fedsn,chen2024gradient}. FL typically employs a parameter server (PS) architecture, where a central server aggregates model updates from participating devices. The efficiency of FL training is heavily dependent on the quality of communication, given the continuous exchange of large volumes of parameters~\cite{wu2024wafbooster}. However, existing FL systems solely rely on terrestrial networks, which are increasingly strained by rising user numbers and service demands~\cite{sacco2021owl, terrigrow,lin2024split}. This often results in delays or congestion in parameter transmission~\cite{yang2023detfed}, significantly affecting FL convergence, especially in remote or underdeveloped regions with poor signal coverage~\cite{10040542}.

\begin{figure}[t!]
\centering
\includegraphics[width=0.5\textwidth]{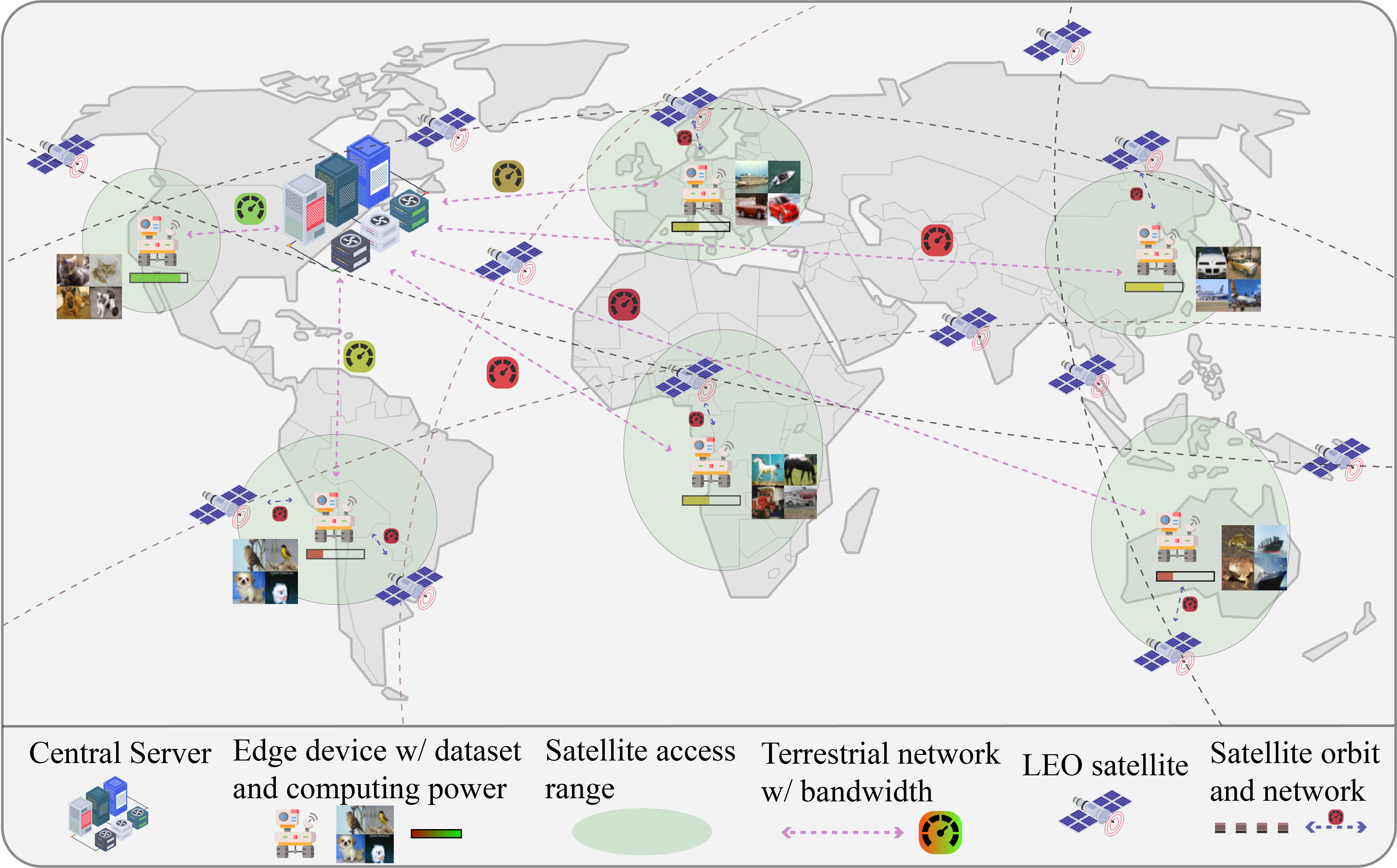}
\caption{Paradigm and challenges in satellite-assisted FL.}
\label{fig_paradigm}
\vspace{-3.5ex}
\end{figure}




Fortunately, the proliferation of low Earth orbit (LEO) satellites is revolutionizing satellite broadband Internet~\cite{zhai2023fedleo, yuan2024satsense,zhang2022enabling,zhao2024leo,yuan2023graph}, providing a crucial solution to the limitations of traditional terrestrial FL. Specifically, SpaceX plans to deploy about $42,000$ LEO satellites for its Starlink constellation~\cite{prevelance_of_satellite2}. LEO satellite networks offer broad coverage, facilitating peer-to-peer communication among numerous edge devices~\cite{10229104}. For example, when devices engaged in FL experience unstable terrestrial connections with central servers, satellite connectivity allows them to rapidly obtain knowledge from multiple peers, thereby improving their local training.

However, despite its promise, FL within satellite-assisted terrestrial networks remains largely unexplored. This is a complex endeavor requiring meticulous design of transmission modes tailored to the intrinsic characteristics of satellite networks and effective utilization of satellite transmissions to address various performance challenges posed by system heterogeneity. Firstly, the limited contact time and bandwidth of satellite networks, particularly in the uplink, significantly constrain the delivery of parameters. Secondly, ground devices typically operate in highly diverse environments, characterized by data heterogeneity, varying terrestrial network conditions, and differing local computing power. These system heterogeneities notably degrade model performance and represent a primary challenge that satellite assistance must effectively address. Fig.~\ref{fig_paradigm} illustrates the satellite-assisted FL paradigm and its main challenges, which we will explore in greater detail in the next section.


To address these challenges, we propose a satellite-assisted FL framework for resource-constrained and heterogeneous environments, named \textit{SatFed}. In SatFed, each edge device trains a personalized model for local inference while also contributing to the global model training. Global model updates occur asynchronously between devices and the server via terrestrial networks, while devices exchange personalized models peer-to-peer through satellite networks for mutual guidance. SatFed utilizes freshness priority queues for parameter transmission over bandwidth-constrained satellite networks, ensuring that devices receive the most up-to-date peer models. Additionally, a multigraph with three edge types is employed to dynamically capture diverse heterogeneities between devices, optimizing the local utilization of peer models received from satellite networks. The three edge types are: i) \textit{similarity edges}, which help devices learn from peers with similar data distributions; ii) \textit{connection edges}, which enable devices to benefit from peers with superior terrestrial bandwidth; and iii) \textit{computation edges}, which monitor differences in device model update speeds and adjust the local learning rate accordingly. Our key contributions can be summarized as follows.


\begin{itemize}
  \item To our knowledge, SatFed represents the pioneering effort to harness LEO satellite networks for enhancing FL within traditional terrestrial networks. It tackles the distinctive transmission characteristics of satellite networks by implementing model freshness queues, effectively mitigating the challenges posed by constrained bandwidth on satellite-assisted performance.
  \item We investigate innovative solutions for addressing system heterogeneity in satellite-assisted terrestrial communication frameworks, employing a multigraph to monitor real-time model transmissions within satellite networks and capture the intricate system heterogeneities. Through meticulous design of local updates, the multigraph-guided integration of satellite-transmitted models markedly improves training performance.
  \item We implement the SatFed prototype using actual LEO satellite networks and datasets. Comprehensive experiments demonstrate its capacity to achieve efficient and robust FL with satellite assistance across various heterogeneous conditions.
\end{itemize}

In the following text, Sec.~\ref{sec:motivation} provides motivation for the design of SatFed by examining the challenges associated with satellite-assisted FL. The system architecture of SatFed is detailed in Sec.~\ref{sec:design}, followed by its implementation and performance evaluation in Sec.~\ref{sec:eval}. Sec.~\ref{sec:relatedwork} explores related work and technical constraints, with Sec.~\ref{sec:conclu} presenting conclusions.


\section{Motivation}\label{sec:motivation}
In this section, we present empirical measurements that underscore the challenges inherent in satellite-assisted FL, thereby strengthening the motivation for the design of SatFed.

\subsection{Constrained Satellite Network Transmission}
\label{subseq::limit_satnet}

\begin{figure}[t]
\centering
\begin{subfigure}{0.24\textwidth}
\includegraphics[width=\textwidth]{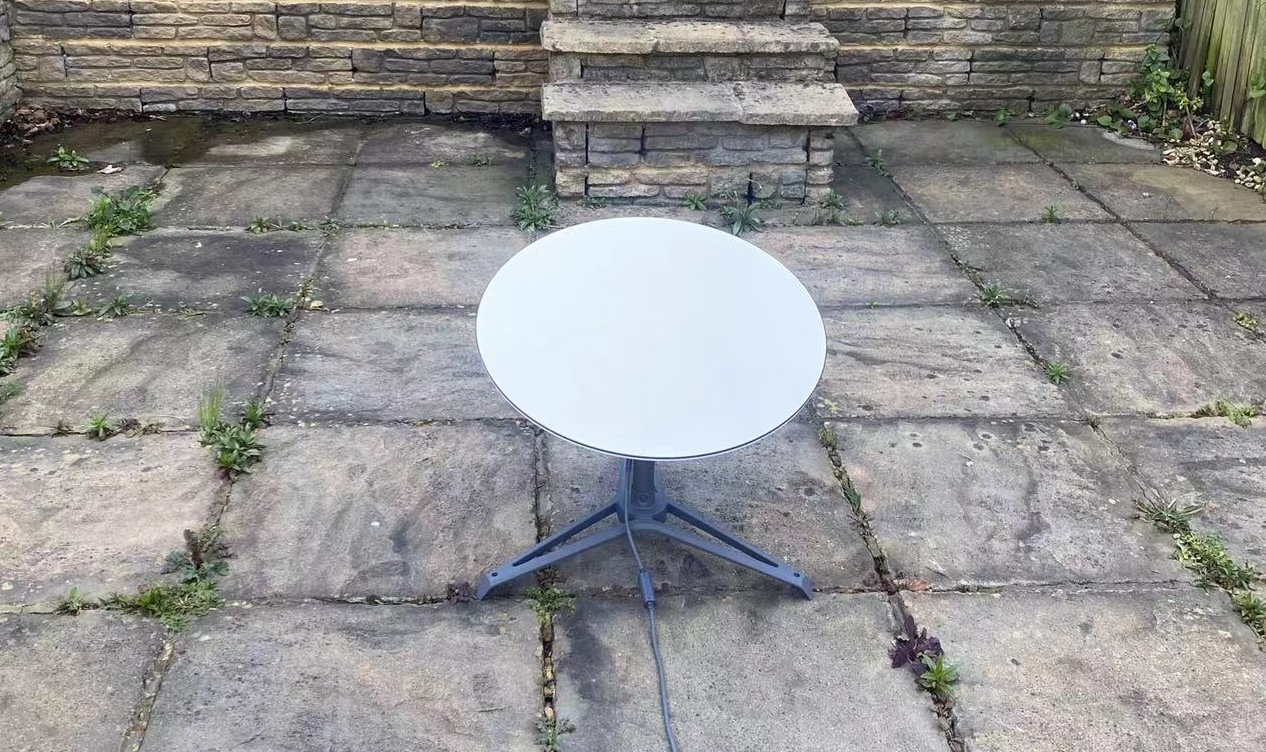}
    \caption{Experimental setup.}
    \label{subfig:motivating_setup}
\end{subfigure}
\hfill
\begin{subfigure}{0.24\textwidth}
\includegraphics[width=\textwidth]{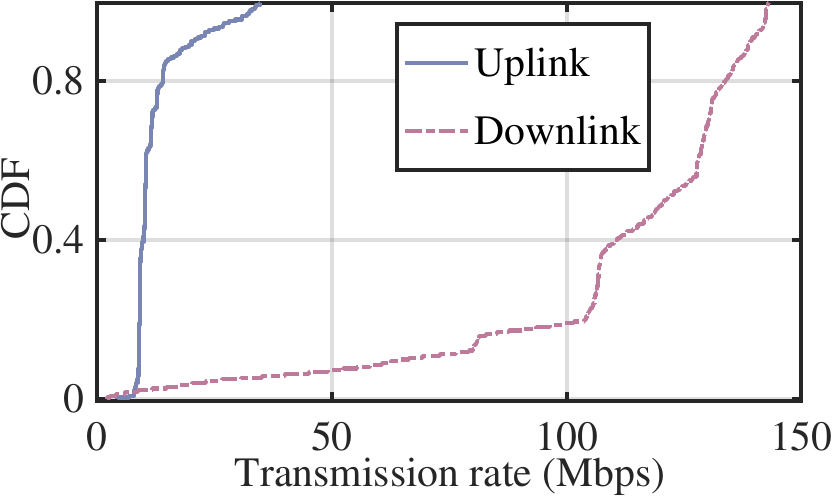}
    \caption{Satellite network speed.}
    \label{subfig:motivating_cdf}
\end{subfigure}
\caption{The uplink in the LEO satellite network represents a significant transmission bottleneck. Fig.~\ref{subfig:motivating_setup} illustrates our experimental setup, while Fig.~\ref{subfig:motivating_cdf} presents the resulting CDF.}
\label{fig:motivating_satband}
\vspace{-3.5ex}
\end{figure}

\begin{figure}[t]
\centering
\begin{subfigure}{0.245\textwidth}
\includegraphics[width=\textwidth]{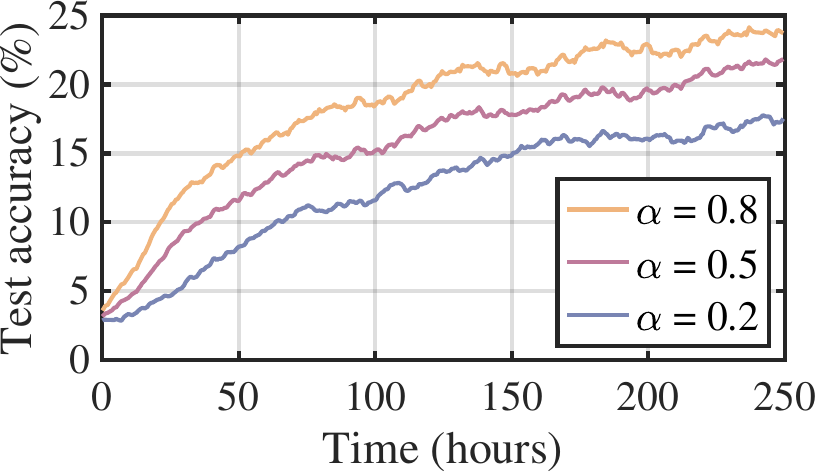}
\caption{Heterogeneity levels.}
\label{subfig:motivating_dataheter1}
\end{subfigure}
\hfill
\begin{subfigure}{0.23\textwidth}
\includegraphics[width=\textwidth]{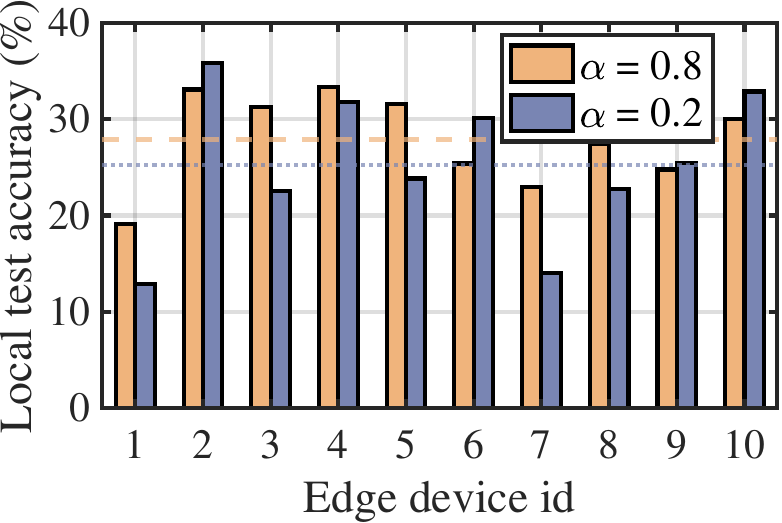}
    \caption{Fairness.}
    \label{subfig:motivating_dataheter2}
\end{subfigure}
\caption{The impact of data heterogeneity. The CIFAR-100 dataset~\cite{CIFAR10} is distributed among 10 edge devices, with local labels following a Dirichlet distribution (a smaller value of $\alpha$ indicates greater heterogeneity)~\cite{chen2022towards}. The experimental parameters are consistent with Sec.~\ref{sec:eval}}
\label{fig:motivating_dataheter}
\vspace{-3.5ex}
\end{figure}

LEO satellites orbit periodically, engaging in bidirectional data transmission when within range of ground devices, primarily constrained by brief contact times and limited uplink bandwidth. For instance, satellites orbiting at an altitude of 700 km complete one orbit approximately every 100 minutes, with contact windows to ground devices lasting only about 10 minutes~\cite{contacttime}. As illustrated in Fig.~\ref{subfig:motivating_setup}, we assess parameter transmission efficiency during these fleeting contact periods using the Starlink LEO satellite communication system. Fig.~\ref{subfig:motivating_cdf} presents the cumulative distribution function (CDF) derived from Iperf~\cite{iperf}, a widely used network measurement tool, revealing an average downlink rate of approximately $100$ Mbps and an uplink rate around $12$ Mbps. For example, downloading a $98$ MB ResNet-50 model~\cite{resnet} from the satellite to each device takes about $8$ seconds, whereas uploading the model back to the satellite requires more than a minute. Thus, the principal bottleneck in satellite network transmission lies in the uplink direction.



In an ideal scenario, each connection between devices and satellites involves extensive model exchanges (as detailed further in Sec.~\ref{subsec:satnetmodel}). However, due to limitations in uplink transmission rates, the successful transmission of models (typically fewer than 10 in our ResNet-50 case) is significantly constrained. Thus, prioritizing critical models for satellite transmission is essential to optimize satellite-assisted performance.


\subsection{Data Heterogeneity}
\label{subsec::data_heter}

The performance of satellite-assisted FL is significantly impacted by data heterogeneity, as devices distributed globally encounter data from diverse contexts~\cite{yang2023efficient}. These devices exhibit non-independent and non-identically distributed (non-IID) data, resulting in varied local optimization objectives~\cite{heter1}. Consequently, model updates from different devices are biased toward their local data distributions. Directly averaging these updates, as is done in the widely used FL method FedAvg~\cite{FedAvg}, often leads to suboptimal performance. To better understand the impact of data heterogeneity on training, we conducted experiments with the ResNet-50 model using FedAvg at varying levels of heterogeneity, as shown in Fig.~\ref{subfig:motivating_dataheter1}. The results clearly indicate that as data heterogeneity increases, performance degrades significantly.


Moreover, relying solely on a single global model can lead to fairness challenges across devices. To investigate this issue, we evaluate each device's local test accuracy under various heterogeneous conditions, as shown in Fig.~\ref{subfig:motivating_dataheter2} (with the dashed line representing the average). It is evident that as heterogeneity increases, the global model exhibits greater performance variance among devices, significantly reducing fairness. Although extensive research has been conducted on the impact of data heterogeneity on FL, most studies focus on PS communication frameworks~\cite{tan2022towards} and do not adapt to emerging hybrid satellite-terrestrial network architectures.



\subsection{Bandwidth Heterogeneity}
\label{subsec::band_heter}

\begin{figure}[t]
\centering

\begin{subfigure}{0.24\textwidth}
\includegraphics[width=\textwidth]{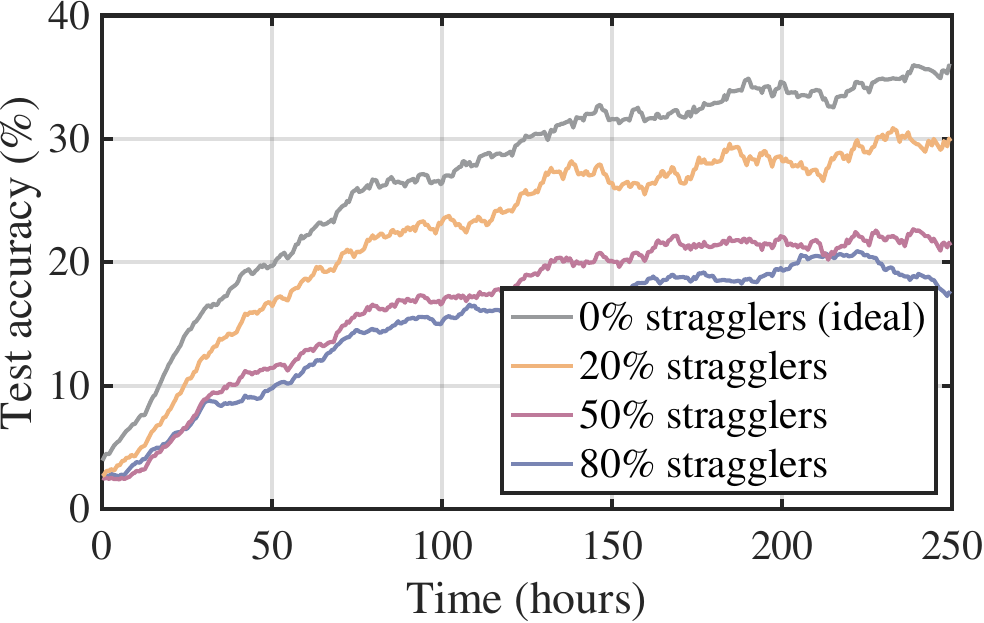}
    \caption{Straggler ratios.}
    \label{subfig:motivating_band1}
\end{subfigure}
\hfill
\begin{subfigure}{0.24\textwidth}
\includegraphics[width=\textwidth]{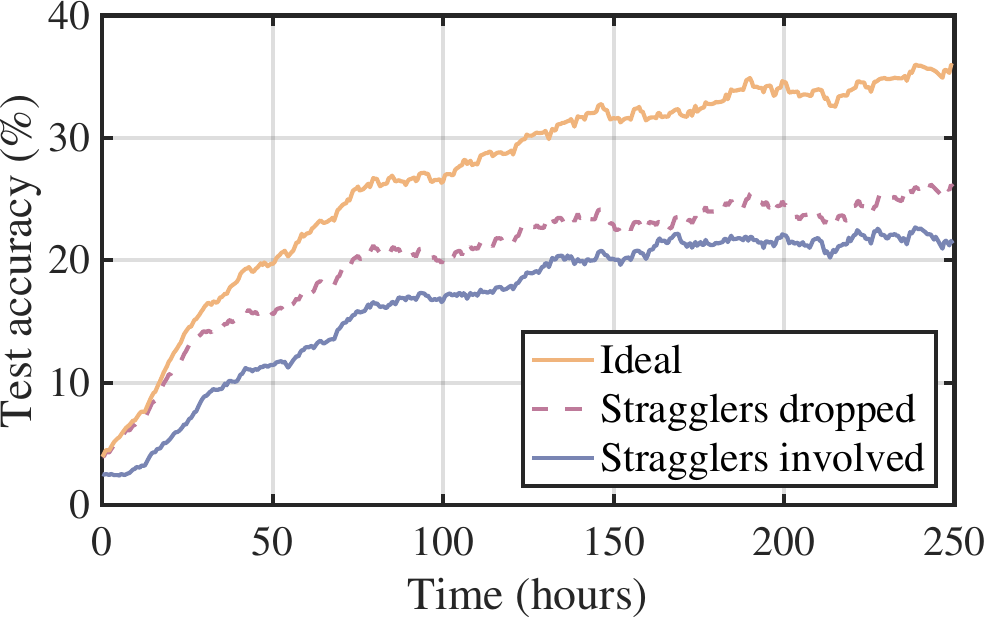}
    \caption{Damage from stragglers.}
    \label{subfig:motivating_band2}
\end{subfigure}

\caption{
The impact of bandwidth stragglers.
\textit{Asynchronously} aggregate local updates to form and evaluate the global model.
Stragglers have a $90\%$ chance of communication blockage or disconnection with the server, retrying every $30$ minutes.
}
\label{fig:motivating_bandheter}
\vspace{-3.5ex}
\end{figure}

Terrestrial networks generally provide higher bandwidth than satellite networks, but they are subject to variability and heterogeneity due to factors such as fluctuating user volumes, routing conditions, and regional infrastructure disparities~\cite{liao2023adaptive}. Devices operating in remote areas or congested environments (a.k.a., stragglers) experience longer delays in parameter transmission , which in turn slows down their local updates and ultimately hampers training performance.



To better understand the impact of bandwidth heterogeneity on FL, we conduct experiments with varying proportions of stragglers. Fig.~\ref{subfig:motivating_band1} shows that as the proportion of stragglers increases, FL experiences significantly slower convergence and a marked deterioration in accuracy. In Fig.~\ref{subfig:motivating_band2}, excluding outdated model updates from aggregation partially alleviates performance degradation but still falls short of the ideal. The comparison highlights that up-to-date models effectively extract local knowledge, while outdated models can negatively impact the system's performance.


Fortunately, the numerous satellites and extensive coverage in LEO constellations can enable edge devices to acquire timely knowledge from multiple peers during periods of ground network congestion or disconnection, thereby mitigating the issue of outdated local updates. However, effective design to address the straggler problem in such a specialized and hybrid network architecture remains largely understudied.



\subsection{Computation Heterogeneity}
\label{subsec::comp_heter}

\begin{figure}[t]
\centering

\begin{subfigure}{0.24\textwidth}
\includegraphics[width=\textwidth]{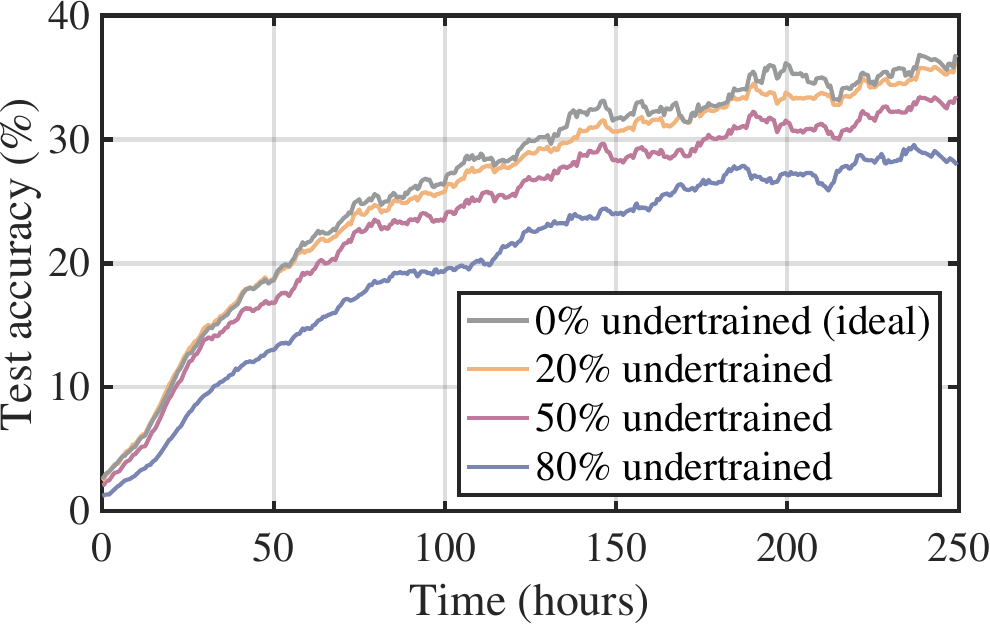}
    \caption{Undertrained ratios.}
    \label{subfig:motivating_comp1}
\end{subfigure}
\hfill
\begin{subfigure}{0.24\textwidth}
\includegraphics[width=\textwidth]{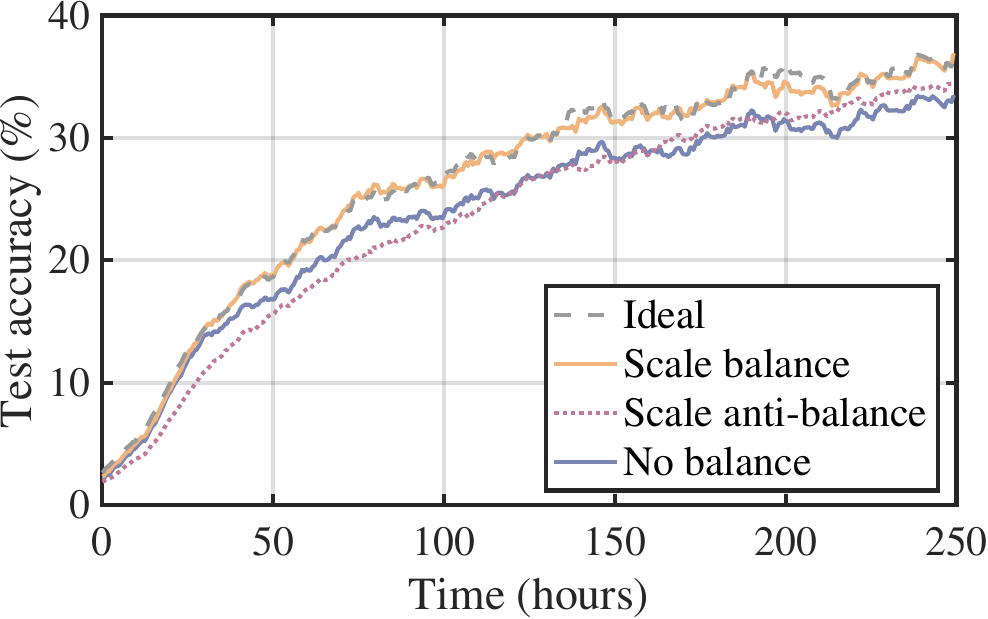}
    \caption{Balance effect.}
    \label{subfig:motivating_comp2}
\end{subfigure}
\caption{The impact of uneven training.
The local update period is set to $30$ minutes, with normal devices having a compute power of $4$ GFLOPS (performance of a Raspberry Pi 4 Model B~\cite{basford2020performance}), while undertrained devices have only $1/5$ of that. The model is ResNet-50 with a batch size of $128$.
Scale balance (scale anti-balance) increases the local learning rates of undertrained (normal) devices by $1.5\times$.}
\label{fig:motivating_compheter}
\vspace{-3.5ex}
\end{figure}

\begin{figure*}[t]
\centering
\includegraphics[width=.9\textwidth]{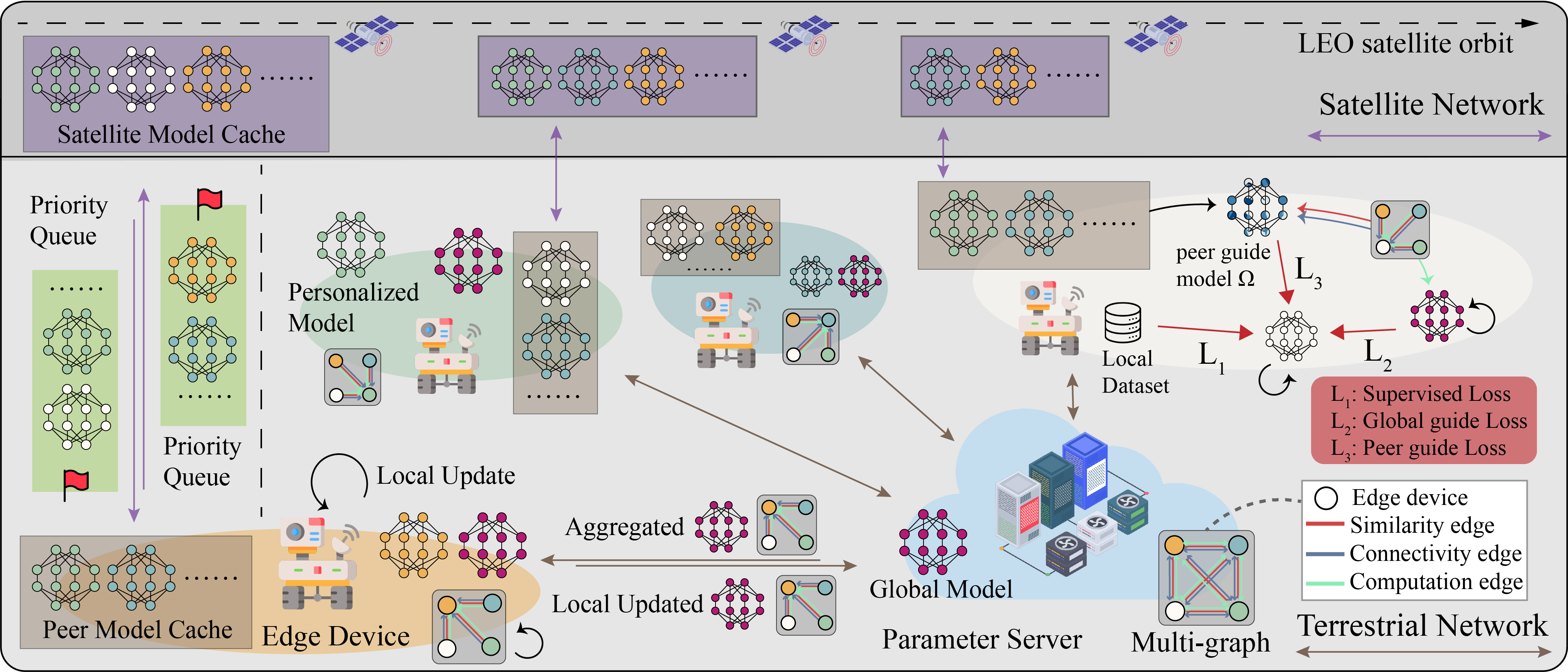}
\caption{
The system overview of SatFed framework.
}
\label{framwork}
\vspace{-3.5ex}
\end{figure*}

{Another} challenge in satellite-assisted FL arises from the difference in computing capabilities among devices~\cite{diao2020heterofl,heter1}.
Certain devices are undertrained due to insufficient computational power within the same local update cycle.
We conduct experiments to investigate how this uneven training in local updates across devices affects FL.
Fig.~\ref{subfig:motivating_comp1} shows that as the number of undertrained devices increases, the test accuracy of the global model significantly decreases.
It should be noted that, as illustrated in Fig.\ref{subfig:motivating_comp2}, the imbalance in the model updates resulting from computation heterogeneity can be mitigated by raising the learning rate for undertrained devices.
Increasing the learning rates for normal devices does not improve performance, highlighting that the improvement in performance comes from balancing update scales.

Motivated by this, the peer-to-peer communication mode of satellite networks is more conducive to edge devices promptly perceiving the model update scales of their peers and accordingly adjusting local learning rates to ensure that all edge devices update with relatively consistent scales.
However, in a hybrid satellite-assisted FL framework, strategies for real-time and precise perception of peer update scales and the corresponding adjustment of local learning rates to achieve optimal performance require careful design.

\section{Framework Design}\label{sec:design}
In this section, we will elaborate on the SatFed framework and its well-designed components.

\subsection{Preliminaries and Overview}

We consider a typical scenario of LEO satellite-assisted FL.
Each edge device $i \in [m] $ possesses its own private dataset $\mathcal{D}_i$ from distribution $ \mathbb{P}_i(x, y)$, where $x$ and $y$ denote the input features and corresponding labels, $m$ is the number of devices.
Devices share a model $f(\omega;\cdot)$ parameterized by weights $\omega^*$. However, as outlined in Section~\ref{subsec::data_heter}, this model cannot effectively accommodate the diverse local data distributions of all devices. Therefore, we utilize personalized device models for local inference, $ \{ f_i(v_i;\cdot) \}_{i \in [m]} $.
The optimization objective follows the state-of-the-art FL methods using personalized models~\cite{ditto}, where all devices are trained within the FedAvg framework to update the global model $\omega^*$ while using it to guide their own personalized model updates:

\begin{equation}
\begin{split}
&\min_{\{ v_i \}} \frac{1}{m} \sum_{i=1}^m  \{ \mathcal{L}_i(v_i) + \frac{\mu}{2} \textrm{Regu}(v_i, \omega^*) \} ,
\end{split}
\end{equation}
\begin{equation}
\textrm{s.t.} \quad \omega^* \in \underset{\omega}{\arg\min} \sum_{i=1}^m  \mathcal{L}_i(\omega),
\end{equation}
where $\mathcal{L}_i(\omega) = \mathbb{E}_{(x,y)\sim \mathcal{D}_i} l(f(\omega;x);y)$ is the empirical loss of edge device $i$, $\mu$ is a hyper-parameter, $N$ denotes the total number of instances over all devices, and $\textrm{Regu}(\cdot,\cdot)$ regulates model difference. 

In SatFed, local updates of the global model are asynchronously transmitted to the server for aggregation via the terrestrial network.
Simultaneously, personalized models are transferred peer-to-peer through the satellite network in a decentralized manner. The personalized model training is guided by both the global model and peers' personalized models.


SatFed employs a model freshness-based priority transmission mode, enabling edge devices to obtain more up-to-date models from peers in bandwidth-limited satellite networks.
In addition, SatFed employs a multigraph that captures multiple heterogeneity relationships in real time using models transmitted in the satellite network.
This in turn aids in distilling knowledge from cached peer models to provide performance and robustness of local updates.
The system overview of SatFed framework is illustrated in Fig~\ref{framwork}.

\subsection{Satellite Network Transmission}
\label{subsec:satnetmodel}
For the transmission of satellite networks, satellite $k$ and device $i$ maintain model caches $\mathcal{C}_{sat}^k$ and $\mathcal{C}^i$ respectively, for storing personalized models from various devices.
As depicted in Fig.~\ref{framwork}, when device $i$ connects to satellite $k$, it uploads models through the satellite network uplink. Satellite $k$ subsequently transmits models via the downlink to device $j$ as it passes overhead.
However, the intricate topology and stringent bandwidth constraints of satellite networks, as discussed in Section~\ref{subseq::limit_satnet}, present significant challenges to naive satellite transmission.


\subsubsection{Drawbacks of Naive Transmission}
\label{subsec:naivetrans}
Given $L$ LEO satellite orbits with multiple satellites operating on each, for each orbit $l \in [L]$, the set $S_l \subseteq [m]$ represents the edge devices it covers.
When two edge devices $i$ and $j$ are simultaneously covered by any orbit, i.e., $\exists l \in [L]$, $ i \in S_l $ and $ j \in S_l $, we term them satellite-network direct, denoted as $SatDir_{i,j}=1$; otherwise, $SatDir_{i,j}=0$.
Notably, any device $i$ may also be simultaneously covered by multiple orbits, denoted as the orbit set $S_{Orb}^{i}$.
It can be inferred that $S_i^{SatDir} = \{j|j \in m, SatDir_{i,j}=1\}$ is the union $ \bigcup \{ S_l| l \in S_{Orb}^{i} \} $.

\begin{figure}[t]
\centering
\begin{subfigure}{0.24\textwidth}
\includegraphics[width=\textwidth]{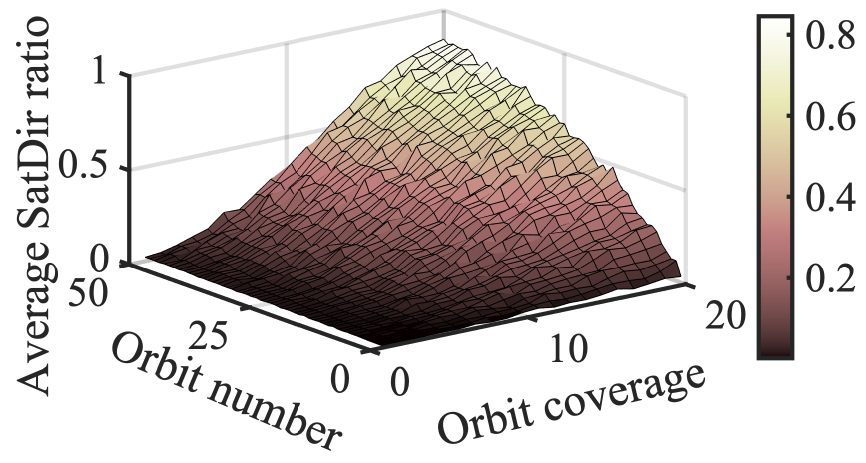}
    \caption{Transmission bias.}
    \label{subfig:transbias}
\end{subfigure}
\hfill
\begin{subfigure}{0.24\textwidth}
\includegraphics[width=\textwidth]{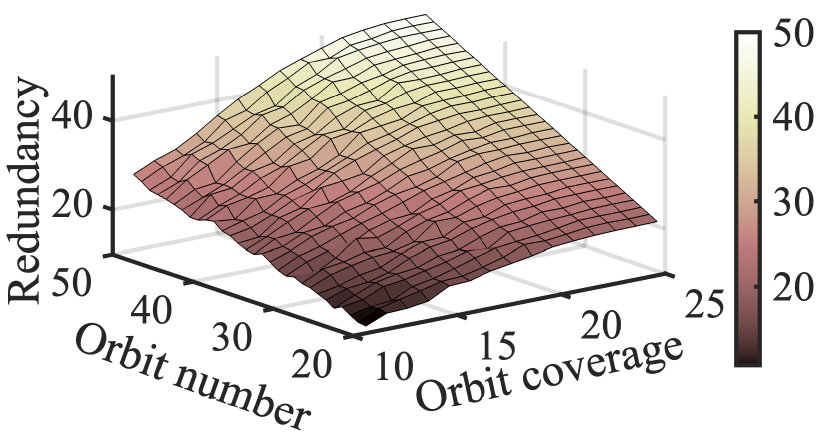}
    \caption{Transmission redundancy.}
    \label{subfig:transredun}
\end{subfigure}

\caption{Simulation of problems with naive satellite network transmission.
Fig.~\ref{subfig:transbias} illustrates the average $ |S_i^{SatDir}| / m$ across different topology configurations, where $m=100$. In Fig~\ref{subfig:transredun}, device $i$ uploads entire $\mathcal{C}_i$ upon connecting to the satellite. Redundancy indicates the ratio of transmitted models to the necessary ones.}
\label{fig:motivating_figures}
\vspace{-3.5ex}
\end{figure}

We consider two naive transmission modes: when device $i$ connects to satellite $k$, it either i) uploads only its personalized model $v_i$ or ii) additionally uploads all cached peer models $\{v_j | v_j \in \mathcal{C}^i \} $.
The former method restricts satellite $k$ in orbit $l$ to caching only $ \{ v_i | i \in  S_l \} $, which limits device $j$ to receiving models solely from devices in $S_j^{SatDir}$, thereby causing transmission bias.
As shown in Fig.~\ref{subfig:transbias}, even with an average LEO satellite orbit covering up to $10$ devices and increasing the number of orbits to $50$, devices may receive fewer than half of the peer models, resulting in notable bias in collaborative knowledge distribution.
The second approach eliminates this bias by allowing satellites to receive models from other orbits. However, it significantly increases the transmission load on the uplink of the satellite network and introduces a substantial amount of redundant transmissions.
When the same model is transmitted to another device through multiple orbits, redundant transmission occurs, consuming the highly limited bandwidth resources of the satellite network.
As shown in Fig.~\ref{subfig:transredun}, as the number of orbits and coverage increases, this redundancy becomes increasingly severe, wasting tens of times the bandwidth resources.

\subsubsection{Model Freshness and Priority Queues}

To tackle transmission bias and redundancy, SatFed implements model priority queues to prioritize the transmission of critical models within the constrained bandwidth of satellite networks.
Initially, we define the \textit{freshness} of each model circulating in the satellite network to indicate its level of obsolescence.
Specifically, each time a device $i$ uploads model $v_i$ to the satellite, it timestamps the model as $v_i^t$, where $t$ represents the upload initiation time.
At any subsequent time $t^*$, the freshness of $v_i^t$ is denoted as $ \mathcal{R}(v_i^t,t^*) = 1 - e^{-\eta(t^* - t)}$, indicating that the closer the upload time of the model $t$ to the current time $t^*$, the higher its freshness, where $\eta$ is a hyperparameter.

Based on freshness, device $i$ constructs upload and download model priority queues, $Q_{i,k}^{U}$ and $Q_{i,k}^{D}$, respectively, when connected to satellite $k$.
At connection establishment time $t^*$, device $i$ compares the freshness $\mathcal{R}(v_j^{t^1},t^)$ and $\mathcal{R}(v_j^{t^2},t^)$ of peer models $v_j^{t^1} \in \mathcal{C}^i$ and $v_j^{t^2} \in \mathcal{C}_{sat}^k$ for all $j \in [m]$: if $\mathcal{R}(v_i^{t^1},t^) > \mathcal{R}(v_i^{t^2},t^*)$, add $v_j^{t^1}$ to the upload queue $Q_{i,k}^{U}$; otherwise, add $v_j^{t^2}$ to the download queue $Q_{i,k}^{D}$.
If a cache $\mathcal{C}$ lacks a model from device $j$, it is treated as $v_j^{-\infty} \in \mathcal{C}$.
The models in the queues are arranged in descending order according to the freshness difference $| \mathcal{R}(v_i^{t^1},t^*) - \mathcal{R}(v_i^{t^2},t^*)|$.
Following the priority sequence in the queues, device $i$ subsequently uploads or downloads models, replacing the corresponding outdated models in the recipient cache.

Thus, SatFed tackles bias and redundancy in satellite network transmission via priority queues.
i) If $v_i$ does not exist in the recipient cache or is extremely outdated, the sender positions $v_i$ at the front of the queue for priority transmission, thereby maximizing the alleviation of transmission bias.
ii) If two different satellites prepare to transmit personalized models from the same device with close upload times to the same recipient, the second model, having minimal freshness difference, is placed at the end of the transmission queue upon receiving the first model. It is transmitted only if bandwidth is significantly idle, thus avoiding redundancy.

\subsection{Heterogeneous multigraph}
\label{subsec:multigraph}

To address the multifaceted heterogeneity issues highlighted in Sec.~\ref{sec:motivation}, SatFed constructs a global multigraph, aiming to leverage model exchanges across satellite networks to capture the diverse relationships between devices, ultimately guiding local updates.
Specifically, we document three types of heterogeneous relationships to construct the multigraph $\mathcal{G}_{m}= ([m], \mathcal{A}_{sim}, \mathcal{A}_{con}, \mathcal{A}_{cmp})$:
i) similarity edges $\mathcal{A}_{sim}$;
ii) connectivity edges $\mathcal{A}_{con}$;
and iii) computation edges $\mathcal{A}_{cmp}$.
The details are as follows.

\subsubsection{Similarity Edges}
\label{subsubsec::simiedges}
As described in Sec~\ref{subsec::data_heter}, diverse devices exhibit unique local empirical risks $\{ \mathcal{L}_i \}_{i \in [m]}$ attributed to differences in training data distribution.
Similarity edges $\mathcal{A}_{sim}$ aim to aid devices in discerning which peer models obtained from the satellite network better align with their optimization goals.
Borrowing from state-of-the-art methods~\cite{pfedgraph}, we measure the similarity of local optimization objectives by calculating the cosine similarity of model parameters during training.
However, in satellite networks where model freshness varies, simple cosine similarity may lead to inaccurate measurements, necessitating freshness calibration.
Therefore, we compute similarity edge $\mathcal{A}_{sim}$ between devices $i$ and $j$ with confidence $\kappa = e^{-|t^1-t^2|} $ as:

\begin{equation}
\begin{split}
\mathcal{A}_{sim}(i, j) = 1 - \frac{v_i^{t^1}\cdot v_j^{t^2}}{||v_i^{t^1}||||v_j^{t^2}||}. \,
\end{split}
\end{equation}
In short, when two models have similar freshness ($\kappa \to 1$), cosine similarity accurately reflects the similarity of their local optimization objectives. Conversely, when their freshness differs significantly ($\kappa \to 0$), cosine similarity fails to reflect their data distribution accurately.
The confidence $\kappa$ adjusts the similarity edge updates, as detailed in Sec.~\ref{subsubsec::multigraphupdate}.


\subsubsection{Connection Edges}
As discussed in Sec.~\ref{subsec::band_heter}, terrestrial bandwidth stragglers result in relatively outdated models. Connection edges help stragglers identify potential helpers.
We monitor the communication frequency $C^i_{S}$ between each edge device $i$ and the parameter server over a specified time window $W$.
Similarly, we record the frequency $C^{i, j}_{D}$ of successful transmissions of personalized models from device $j$ to $i$ via the satellite networks.
Subsequently, the connection edge is denoted as:

\begin{equation}
\begin{split}
\mathcal{A}_{con}(i, j) = ( C^j_{S} + \lambda \times C^{i, j}_{D} ) / C^i_{S} ,
\end{split}
\label{Eq:Acon}
\end{equation}
where $\lambda > 0$ is the weight balancing the two indicators.
$\mathcal{A}_{con}(i, j)$ is large if device $j$ has significantly more exchanges with the server than device $i$ in a short period of time ($ C^j_{S} \gg C^i_{S}$).
This suggests that the $v_i$ can rely more on peer model $v_j$ to keep up with the global training pace.
Furthermore, $C^{i, j}_{D}$ encourages $v_i$ to learn from more familiar and easily accessible peer models, thereby reducing fluctuations in the convergence of the entire system.
Furthermore, $C^{i, j}_{D}$ encourages local updates to learn more from familiar and easily accessible models of the same type, thereby reducing convergence fluctuations.

\subsubsection{Computation Edges}
\label{subsubsec::cmpedges}
To mitigate the global model degradation caused by the uneven training mentioned in Sec.~\ref{subsec::comp_heter}, we devise computation edges $\mathcal{A}_{cmp}$ to track the differences in update speeds across devices, thereby assisting in compensating for undertrained ones.
Specifically, we define the update speed of $vi$ as the Euclidean distance between parameters before and after the update, divided by the time interval:

\begin{equation}
\begin{split}
\mathcal{S}_i =  \frac{ || v_j^{t} - v_j^{t^{\prime}} || }{t - t^{\prime}}  .
\end{split}
\label{Eq:Si}
\end{equation}
Subsequently, $\mathcal{A}_{cmp} $ is computed as the ratio of model update speeds between different devices:

\begin{equation}
\begin{split}
\mathcal{A}_{cmp}(i, j) = \mathcal{S}_j / \mathcal{S}_i .
\end{split}
\label{Eq:Acmp}
\end{equation}
Based on $\mathcal{A}_{cmp}$, SatFed evaluates and compares the model update speeds of all edge devices, and can adjust the local learning rate of undertrained models to prevent detrimental impacts on global model aggregation.

\subsubsection{multigraph Update}
\label{subsubsec::multigraphupdate}
The updates of the multigraph $\mathcal{G}_{m}$ are handled by both the server and the devices, and synchronized at the server side.
At device $i$, updates occur at time $t^*$ when downloading any peer model $v_j^t$ via the satellite network, replacing the locally cached old model \( v_j^{t^{\prime}} \).
$\mathcal{A}_{sim}$ updates using weights based on the confidence mentioned earlier:

\begin{equation}
\begin{split}
\mathcal{A}_{sim}(i, j) = \kappa \times \frac{v_i^{t^*} \cdot v_j^t}{||v_i^{t^*}||||v_j^t||} + ( 1 - \kappa) \mathcal{A}_{sim}(i, j),
\end{split}
\end{equation}
where confidence $\kappa =  e^{-|t^*-t|}$.
Additionally, device $i$ recalculates the count of times it received $v_j$ within the time window $W$ to update $C^{i, j}_{D}$, and subsequently adjusts $\mathcal{A}_{con}(i, j)$ using Eq.~\ref{Eq:Acon}.
Finally, compute $\mathcal{S}_j = \frac{ || v_j^{t} - v_j^{t^{\prime}} || }{t - t^{\prime}}$, and similarly calculate $S_i$, updating $\mathcal{A}_{cmp}(i, j)$ using Eq.~\ref{Eq:Acmp}.

As shown in Fig.~\ref{framwork}, the update of $\mathcal{G}_{m}$ at device $i$ is transmitted to the server along with the local update of the global model $\omega$.
Each time a local update is received from device $i$, the server updates $C^i_{S}$ and correspondingly recalculates $\mathcal{A}_{con}$ using Eq.~\ref{Eq:Acon}.

\subsection{Model Training}
This section will elucidate how SatFed leverages multigraph $\mathcal{G}_{m}$ to guide devices in extracting beneficial knowledge from transmitted models in satellite networks, thereby enhancing local updates and improving training performance.

\subsubsection{Peer Guide Model}
\label{subsubsec:peerguidemodel}
In SatFed, devices cache many peer models from the satellite network, but these models differ greatly in their usefulness for local updates.
Thus, using $\mathcal{G}_m$, we calculate the weight $\mathcal{T}(i, j)$, representing device $i$'s dependency on model $v_j$ for local updates.
The dependency conforms to the two principles, consistent with the design goals of $\mathcal{A}_{sim}$ and $\mathcal{A}_{con}$:
i) the more similar their local optimization objectives, the stronger the mutual dependency;
ii) the more frequent the recent communication between device $j$ and the server, and the more often device $j$'s model is sent to device $i$, the stronger $i$'s dependency on $j$.
Therefore, $\mathcal{T}(i, j)$ is defined as follows:

\begin{equation}
\begin{split}
\mathcal{T}(i, j) =   \mathcal{A}_{sim}(i, j) +  \alpha \times \mathcal{A}_{con}(i, j) ,
\end{split}
\end{equation}
where $\alpha$ is a balancing factor.

Using all peer model $v_j^{t^{\prime}}$ in cache $\mathcal{C}_i$, combined with the respective dependency $\mathcal{T}(i, j)$, we generate a peer-guided model $\Omega_i$ for device $i$:


\begin{equation}
\begin{split}
\Omega_i = \frac{1}{|\mathcal{C}^i|} \sum_{ v_j^{t^{\prime}} \in  \mathcal{C}^i } 
e^{t-{t^{\prime}}}  \times \mathcal{T}(i, j)  \times v_j^{t^{\prime}}  ,
\end{split}
\end{equation}
where $|\mathcal{C}^i|$ is the cache size, $t$ is the current time, and the exponential term aims to diminish the impact of outdated models in the cache.
In short, $\Omega_i$ represents a synthesis of peer models that are ahead in global update progress and closely aligned with the optimization goals of $v_i$.

\subsubsection{Global Model Update}
\label{subsubsec:lradjust}
In SatFed, the global model $\omega^{*}$ is locally trained by devices, with model parameters transmitted via the terrestrial network to a parameter server for asynchronous updates.
However, naive local updates result in uneven training mentioned in Sec~\cref{subsec::comp_heter}, significantly affecting model performance. We address this by using $\mathcal{A}_{cmp}$ to dynamically adjust the local learning rate based on the real-time peer model update speeds.
Specifically, we use $U^i =\sum_{j \in [m]} \mathcal{A}_{cmp}(i, j) / m$ to denote the average update scale of peers for each device $i$.
If $ U^i \gg 1$, it indicates that device $i$ is globally lagging behind and its learning rate needs to be increased appropriately for better global model aggregation.

Thus, SatFed establishes dynamic adaptive learning rate for device $i$:
\begin{equation}
\begin{split}
\eta^i =  \eta \times ( 1 + \gamma \times \ln{\max(1, U^i) } ),
\end{split}
\end{equation}
where $\gamma$ represents the gain coefficient.
If the updating speed of $v_i$ outpaces the global ($U_i \in (0,1)$), its learning rate $\eta^i$ remains hyperparameter $\eta$; otherwise, $\eta^i$ is logarithmically increased with $U_i$.
After each round of local updates begins, device $i$ receives the latest global model $\omega^*$ from the server.
It uses a learning rate $\eta^i$ to train for $R$ rounds of gradient descent on its dataset $\mathcal{D}_i$, obtaining updated parameters $\omega^{*}_i$.
The updated parameters are then sent back to the server for asynchronous aggregation with a fixed weight $\beta$.

\begin{algorithm}[tb]
\caption{SatFed}
\label{alg:SatFed}

\textbf{Input}: learning rate $ \eta $, number of local epochs $R$, hyper-parameters $\beta$, $\mu$ and $\lambda$\\
\textbf{Output}: $ \{ v_i \}_{i \in [m]} $\\
\textbf{Initialization}: $\mathcal{G}_{m}$, $ \{ v_i \}_{i \in [m]} $, $\omega^{*}$\\
\textbf{Server executes}:
\begin{algorithmic}[1] 
\FOR{each device $i$ $\in$ $[m]$ \textbf{asynchronously}}
\STATE Server sends $\omega^{*}$, $\mathcal{G}_{m}$ to device $i$ \hfill\algorithmiccomment{\textit{terrestrial}} \\
\STATE Device $i$ updates $ \mathcal{C}^i $, $\mathcal{G}_{m}$ \hfill\algorithmiccomment{\textit{satellite}}\\ 
\STATE $\omega_i^{*} \gets $ LocalUpdate ($ v_i $, $ \omega^{*} $, $ \mathcal{C}^i $, $\mathcal{G}_{m}$, $\mu$, $\lambda$, $\eta$)
\STATE Device $i$ sends $ \omega_i^{*} $, $\mathcal{G}_{m}$ back\hfill\algorithmiccomment{\textit{terrestrial}}
\STATE $\omega^{*} = \beta \times \omega_i^{*} + ( 1 - \beta) \times \omega^{*} $
\STATE Server updates $\mathcal{G}_{m}$ based on Sec.~\ref{subsubsec::multigraphupdate}
\ENDFOR
\end{algorithmic}
\textbf{}

\textbf{LocalUpdate ($ v_i $, $ \omega^{*} $, $ \mathcal{C}^i $, $\mathcal{G}_{m}$, $\mu$, $\lambda$, $\eta$)}:

\begin{algorithmic}[1]
\STATE Build $\Omega_i$ based on Sec.~\ref{subsubsec:peerguidemodel}
\STATE Compute $\eta^i$ based on Sec.~\ref{subsubsec:lradjust}
\STATE $\omega_i^{*} = \omega^{*}$
\FOR{each local epoch $r$ = 1,2,...,$R$}
\STATE Randomly selects a batch $\mathcal{B}_i$ from $\mathcal{D}_i$
\STATE  $\omega_i^{*} = \omega_i^{*} - \eta^i\nabla l_i(\omega_i^*;\mathcal{B}_i) $

\STATE  $v_i = v_i - \eta\nabla l_i(v_i;\mathcal{B}_i) -\eta\mu(v_i - \omega^{*}_{i}) - \eta\lambda(v_i - \Omega_i)$
\ENDFOR
\STATE \textbf{return} $\omega_i^{*}$
\end{algorithmic}
\end{algorithm}

\subsubsection{Personalized Models Update}

The training of the personalized model $v_i$ is guided by both the global model $\omega^{*}$ and the peer guide model $\Omega_i$.
Note that, due to satellite assistance, local updates of $v_i$ no longer need to wait for $\omega^{*}$ transmission and can proceed continuously.
The local objective for the personalized model $v_i$ is as follows:

\begin{equation}
\begin{split}
\min_{\{ v_i \}} \frac{1}{m}\sum_{i=1}^m  \{ \underbrace{\mathcal{L}_i(v_i)}_{L_{\text{sup}}} + \frac{\mu}{2}\underbrace{|| v_i - \omega^* ||^2_2}_{L_{\text{Terrestrial}}} + \frac{\lambda}{2} \underbrace{ || v_i - \Omega_i||^2_2}_{L_{\text{Satellite}}}\} ,
\end{split}
\end{equation}
where the optimization objective consists of three components:
local supervised learning, global model guidance from the terrestrial network, and peer model guidance from the satellite network, where $\mu$ and $\lambda$ are both regularization strength.

The pseudocode details of SatFed are provided in Algorithm~\ref{alg:SatFed}.
Note that while SatFed is tailored for terrestrial and satellite networks, it offers insights into hybrid centralized/decentralized FL frameworks in general.

\section{Performance Evaluation}
\label{sec:eval}
This section presents numerical results to evaluate the training performance of the proposed SatFed framework and the effectiveness of each individual component.

\subsection{Implementation}

We implement SatFed using Python 3.7 and PyTorch 1.12.1 and train it with the NVIDIA GeForce RTX 3090 GPUs.
In the experiments, $N=20$ edge devices are deployed, along with $K=20$ satellites orbiting the Earth, distributed across $J=10$ LEO orbits.
We use two widely acclaimed image datasets image datasets: Fashion-MNIST \cite{xiao2017fashion} and CIFAR-100 \cite{CIFAR10}, using a Dirichlet distribution to generate non-IID label distributions for devices~\cite{chen2022towards}, and set the $\alpha$ parameter to $0.2$ for strongly non-IID settings and $0.5$ for relatively IID settings.
Data quantities per device are randomly assigned, ranging from $1500$ to $2000$.
We employe a ResNet-50 model of $98$ MB size for CIFAR-100 (ResNet-18 for Fashion-MNIST).
The Adam optimizer~\cite{kingma2014adam} is used, with the learning rate $\eta$ set to $0.001$ and the batch size set to $128$.
The loss function $\mathcal{L}_i$ is the cross-entropy loss.

According to real LEO satellite networks~\cite{contacttime}, each satellite orbit has an operational cycle of $100$ minutes and covers approximately $3$ edge devices, with a contact time of about $10$ minutes per device.
The bandwidth of the satellite network is set to $100$ Mbps for downlink and 10 Mbps for uplink as measured.
The local update period is set to $30$ minutes. Half of the devices have a compute power of $4$ GFLOPS (equivalent to a Raspberry Pi 4 Model B), while the others have only $1/5$ of that.
Half of the devices typically operate under ideal $200$ Mbps terrestrial network bandwidth, while the other half have a $90\%$ chance of communication blockage or disconnecting from the server, retrying every $30$ minutes.


\subsection{Overall Performance}

In this section, we evaluate the overall performance of the SatFed framework in terms of test accuracy and convergence speed.
In addition, we assess the impact of satellite network transmission on SatFed's training.
To investigate the advantages of SatFed, we compare it with five other benchmarks:

\begin{itemize}
  \item \textbf{FedAvg:} The server waits for all devices to return their parameter updates, aggregates them into a new global model, and sends it to devices to start the next round.~\cite{FedAvg}.
  \item \textbf{FedAsync:} The updated models from all devices are aggregated into the latest global model in a asynchronous manner~\cite{asynchronousFL}.
  \item \textbf{Ditto:} Each device maintains a personalized model, trains a global model using FedAvg, and incorporates it to aid personalized model updates with L2 regularization~\cite{ditto}.
  \item \textbf{Ditto-Async:} The asynchronous variant of Ditto updates the global model following FedAsync.
  \item \textbf{SatFed(--):} Employing SatFed’s continuous local update method without incorporating peer guide model during training.
\end{itemize}

\subsubsection{The Overall Performance of SatFed}

\begin{table}[t]
\centering
  \begin{tabular}{lSS}
  \toprule
    \multirow{2}{*}{\textbf{Method}} &
    \multicolumn{2}{c}{\textbf{Fashion-MNIST} (\%)} \\
    & {$\alpha=0.2$ (Non-IID)} & {$\alpha=0.5$ (IID)} \\
      \midrule
    {FedAvg} & {91.75$\pm$5.07} & {91.26$\pm$4.65} \\ 
    {FedAsync} & {92.48$\pm$4.67 (5.66$\times$)} & {92.07$\pm$3.40 (3.39$\times$)} \\
    {Ditto} & {94.07$\pm$4.50 (1.08$\times$)} & {93.41$\pm$3.87 (0.89$\times$)} \\
    {Ditto-Async} & {93.88$\pm$4.94 (3.72$\times$)} & {92.85$\pm$3.47 (2.36$\times$)} \\
    {SatFed(--)} & {91.89$\pm$7.08 (3.34$\times$)} & {92.05$\pm$3.77 (3.70$\times$)}  \\
    {\textbf{SatFed}} & {\textbf{95.08}$\pm$3.02 (\textbf{13.82}$\times$)} & {\textbf{94.48}$\pm$2.23 (\textbf{12.16}$\times$)} \\
    \bottomrule
    \toprule
    \multirow{2}{*}{\textbf{Method}} &
    \multicolumn{2}{c}{\textbf{CIFAR-100} (\%)} \\
    & {$\alpha=0.2$ (Non-IID)} & {$\alpha=0.5$ (IID)} \\
      \midrule
    {FedAvg} & {50.30$\pm$4.39} & {54.57$\pm$2.91} \\ 
    {FedAsync} & {48.34$\pm$5.91 (12.89$\times$)} & {54.20$\pm$3.27 (12.18$\times$)} \\
    {Ditto} & {58.56$\pm$5.16 (3.39$\times$)} & {52.82$\pm$3.17 (1.31$\times$)} \\
    {Ditto-Async} & {56.31$\pm$7.09 (22.22$\times$)} & {53.06$\pm$4.32 (10.91$\times$)} \\
    {SatFed(--)} & {58.41$\pm$8.14 (25.72$\times$)} & {53.75$\pm$3.92 (11.84$\times$)}  \\
    {\textbf{SatFed}} & {\textbf{60.82}$\pm$5.23 (\textbf{100.60}$\times$)} & {\textbf{56.92}$\pm$2.97 (\textbf{19.19}$\times$)}\\
    \bottomrule
  \end{tabular}
  \caption{The converged test accuracy (with device standard deviation) and speed of SatFed and other benchmarks, highlighting the optimal in bold. The speed is the reciprocal of the ratio between the epochs needed to reach FedAvg's converged accuracy and the epochs required by FedAvg itself.}
  \label{table:overall_perform}
  \vspace{-3.5ex}
\end{table}

\begin{figure}[t]
\centering
\begin{subfigure}{0.24\textwidth}
\includegraphics[width=\textwidth]{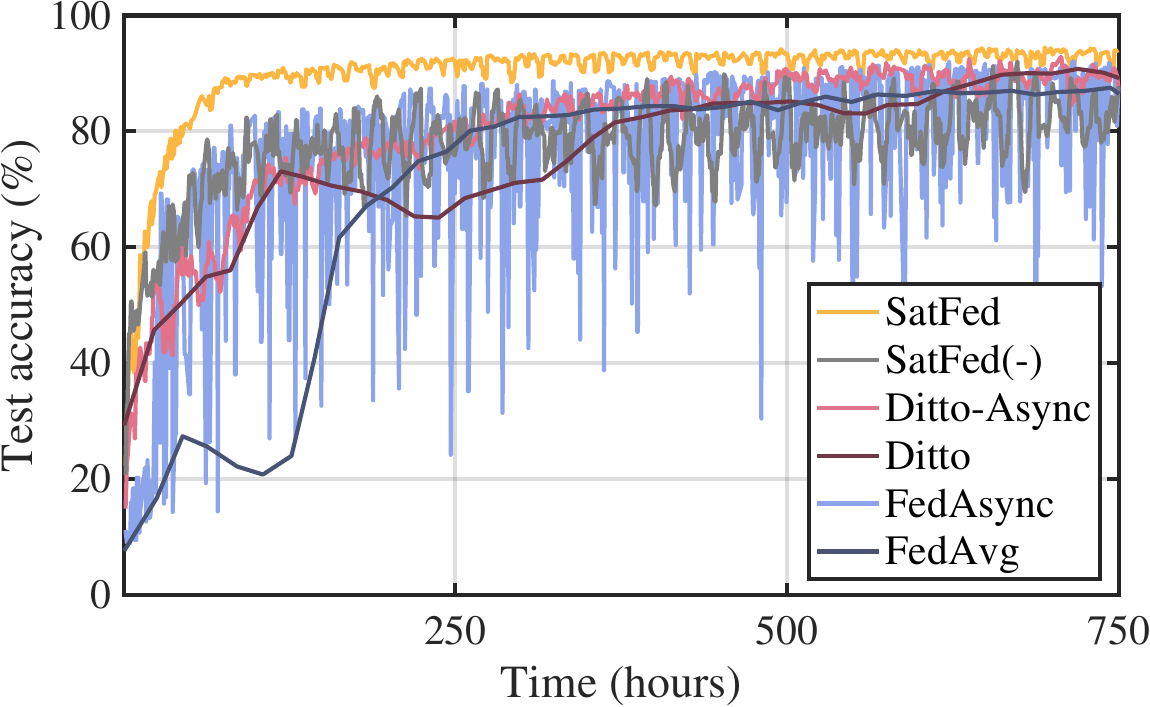}
    \caption{Fashion-MNIST (IID).}
    \label{subfig:overall_curve_gbs_iid}
\end{subfigure}
\hfill
\begin{subfigure}{0.24\textwidth}
\includegraphics[width=\textwidth]{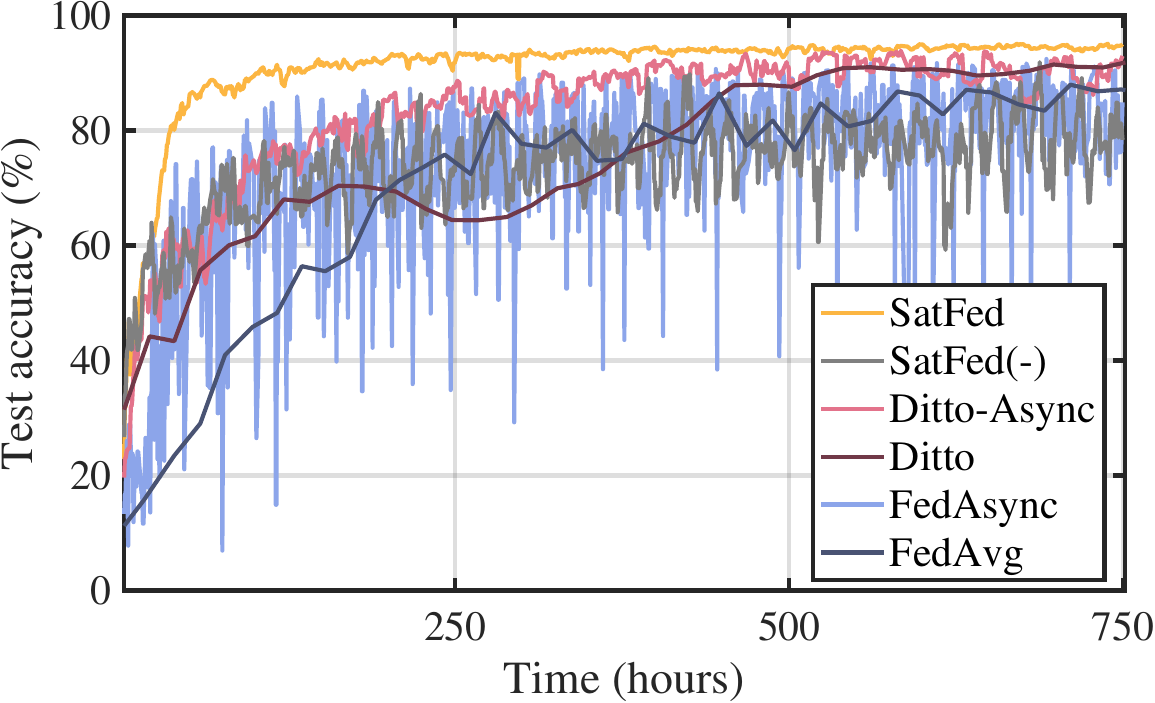}
    \caption{Fashion-MNIST (non-IID).}
    \label{subfig:overall_curve_gbs_noniid}
\end{subfigure}
\hfill
\begin{subfigure}{0.24\textwidth}
\includegraphics[width=\textwidth]{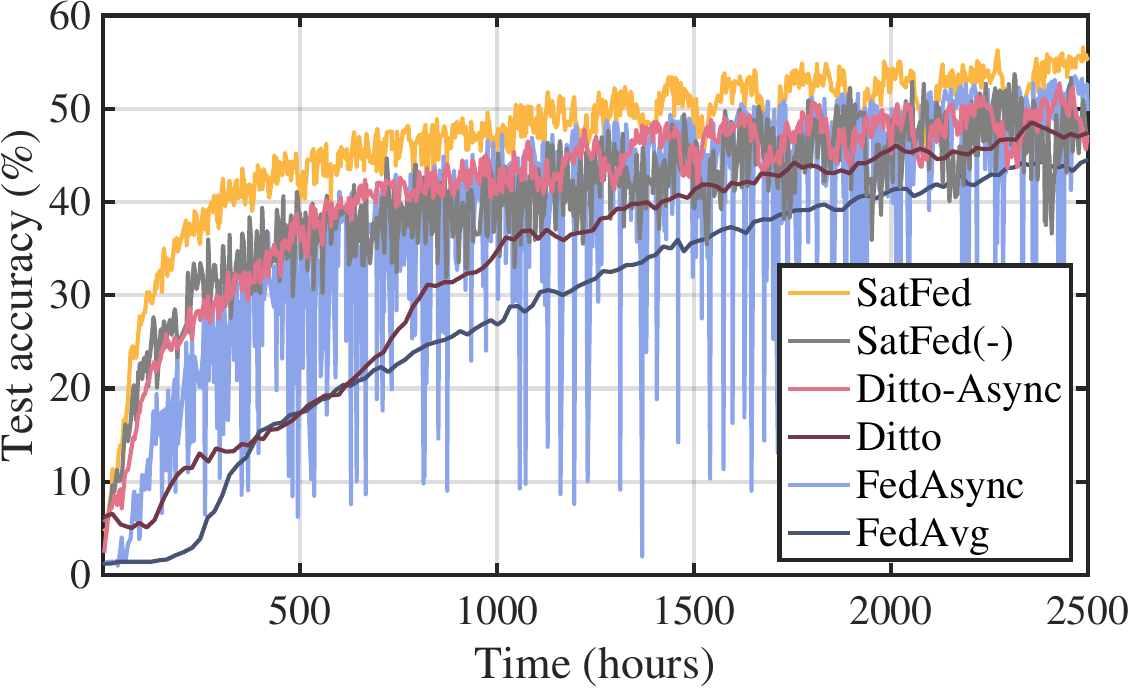}
    \caption{CIFAR-100 (IID).}
    \label{subfig:overall_curve_eus_iid}
\end{subfigure}
\hfill
\begin{subfigure}{0.24\textwidth}
\includegraphics[width=\textwidth]{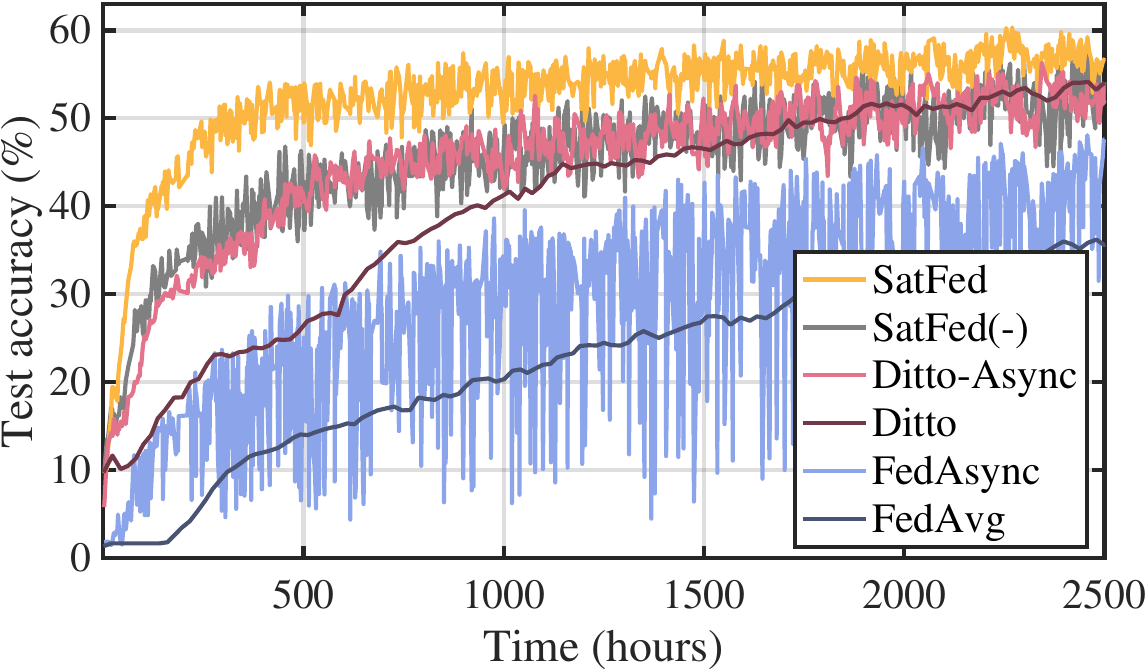}
    \caption{CIFAR-100 (non-IID).}
    \label{subfig:overall_curve_eus_noniid}
\end{subfigure}
\caption{The test accuracy for Fashion-MNIST and CIFAR-100 dataset under IID and non-IID settings.}
\label{fig:simulation_overall_curve}
\vspace{-3.5ex}
\end{figure}

\begin{figure}[t]
\centering
\begin{subfigure}{0.24\textwidth}
\includegraphics[width=\textwidth]{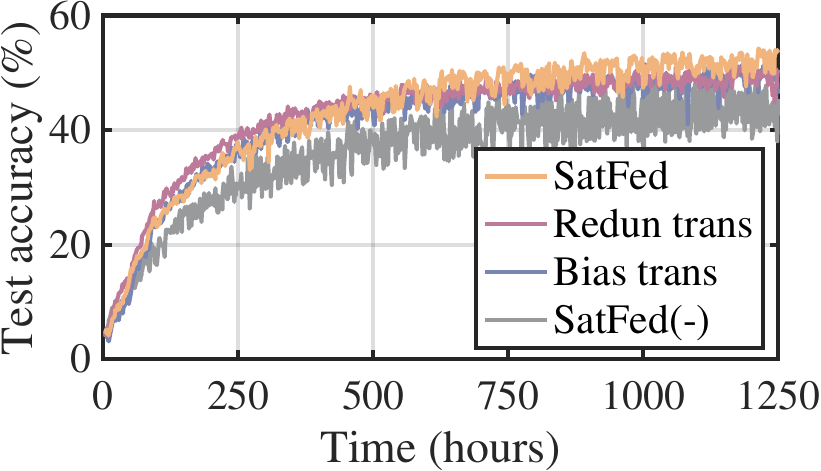}
    \caption{CIFAR-100 (IID).}
    \label{subfig:satel_trans_impact_iid}
\end{subfigure}
\hfill
\begin{subfigure}{0.24\textwidth}
\includegraphics[width=\textwidth]{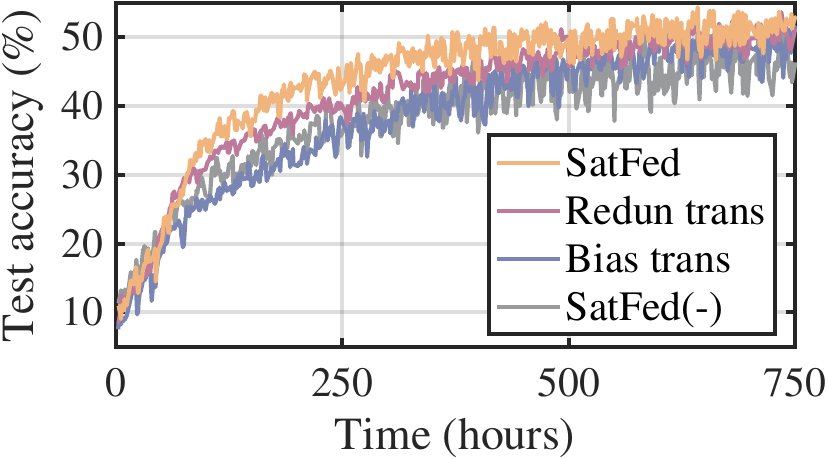}
    \caption{CIFAR-100 (non-IID).}
    \label{subfig:satel_trans_impact_noniid}
\end{subfigure}
\caption{The impact of satellite network transmission methods on training performance. Bias and redundant transmissions are detailed in Section~\ref{subsec:naivetrans}, while SatFed employs the freshness-priority queue scheme.}
\label{fig:simulation_satel_trans_impact}
\vspace{-3.5ex}
\end{figure}

Fig.~\ref{fig:simulation_overall_curve} presents the test accuracy of SatFed compared to other benchmarks on the Fashion-MNIST and CIFAR-100 datasets.
{FedAvg and Ditto} exhibit significantly slower convergence due to waiting for bandwidth-straggling devices during each global model aggregation.
In contrast, FedAsync and Ditto-Async accelerate convergence at the cost of training instability, attributed to the degradation caused by outdated models from asynchronous updates.
It is evident that SatFed achieves the fastest convergence among all benchmarks, facilitated by its connection edges, enabling stragglers to learn from high-bandwidth peers via satellite networks, thereby alleviating local model drag on system convergence progress.
Moreover, SatFed achieved significantly higher test accuracy than other benchmarks, thanks to similarity edges that enables devices to selectively collaborate with peers having similar data distributions.
Besides, computation edges adpat to differences in model update speeds, guiding the adjustment of local learning rates to mitigate the negative impact of uneven training on the global model.
In addition, SatFed(-) did not achieve the same performance as SatFed, suggesting that peer guidance is the key to SatFed's superior performance.


\begin{figure}[t]
\centering
\begin{subfigure}{0.24\textwidth}
\includegraphics[width=\textwidth]{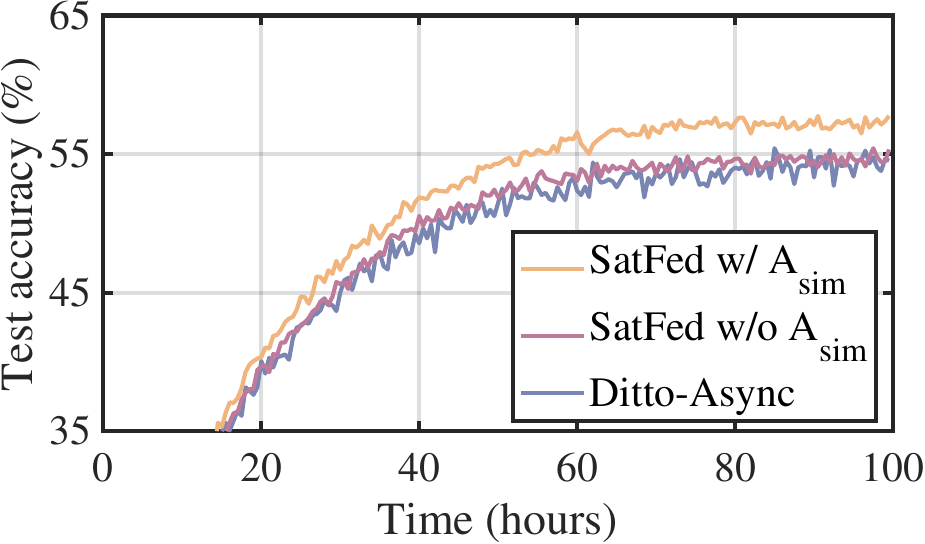}
    \caption{CIFAR-100 (IID).}
    \label{fig:motivating_stragglers}
\end{subfigure}
\hfill
\begin{subfigure}{0.24\textwidth}
\includegraphics[width=\textwidth]{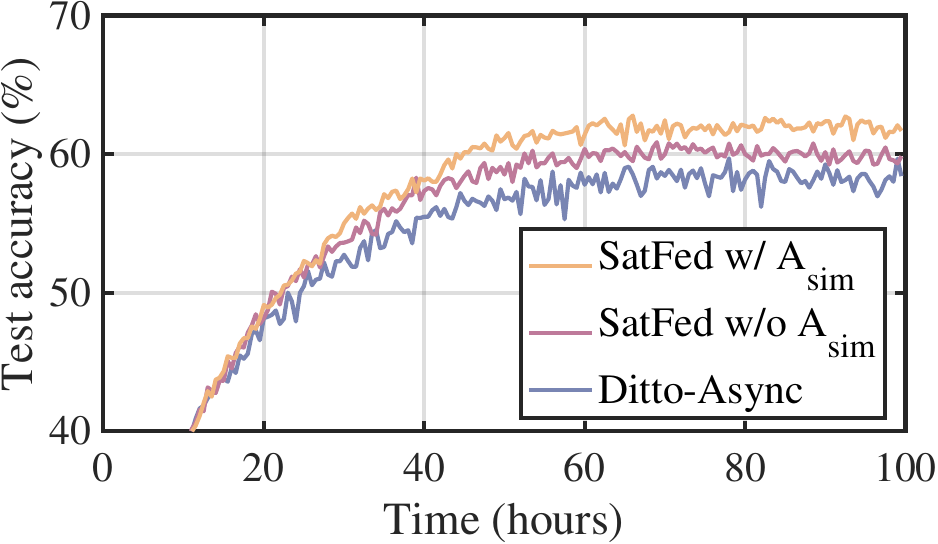}
    \caption{CIFAR-100 (Non-IID).}
    \label{fig:motivating_stragglers}
\end{subfigure}

\caption{Ablation experiments for similarity edges. To emphasize the impact of data heterogeneity, all edge devices are under ideal bandwidth and computational conditions.}
\label{fig:ablation_simi}
\vspace{-3.5ex}
\end{figure}

Table~\ref{table:overall_perform} presents the converged accuracy and speed for Fashion-MNIST and CIFAR-100 datasets.
It's evident that SatFed achieves $2.35\%$ and $2.26\%$ ($1.07\%$ and $1.01\%$) higher test accuracy in IID and non-IID settings on Cifar-100 (Fashion-MNIST) compared to other methods, with convergence speeds up to $100\times$ faster than FedAvg.
SatFed's fast convergence is mainly due to its satellite-assisted guidance from high-bandwidth devices, preventing the personalized models from becoming outdated.
Compared to FedAvg (FedAsync), Ditto (Ditto-Async) achieves better test accuracy in non-IID environments, with an average improvement of $8.12\%$ due to the use of personalized models.
Compared to Ditto-Async, SatFed effectively enhance collaboration among similar peers, while also sensing and compensating for uneven training, further improving performance.




\subsubsection{The Impact of Satellite Network Transmission}

Fig.~\ref{fig:simulation_satel_trans_impact} shows the impact of different satellite network transmission modes on SatFed's training performance.
As described in Sec.~\ref{subsec:naivetrans}, {if} devices upload only their own models, they can only receive models from their satellite-network direct peers, leading to biases that undermine the effectiveness of peer-guided models.
In contrast, the redundant transmission mode involves devices uploading all cached models to satellites, thereby mitigating bias in peer-guided model knowledge sources.
However, this results in a surge in transmission loads, overwhelming the highly constrained uplink links, causing massive model upload failures.
It's evident that SatFed using a freshness-priority model queue achieves the best performance.
This is because it prioritizes the transmission of more critical models based on freshness in constrained satellite networks, ensuring the overall model currency throughout the satellite networks. Consequently, peer-guidance knowledge remains up-to-date, resulting in training performance improvements.

\subsection{Micro-benchmark}


\subsubsection{Similarity Edges}

Fig.~\ref{fig:ablation_simi} illustrates the effect of similarity edges on the training of the CIFAR-100 dataset.
Clearly, SatFed with $\mathcal{A}_{sim}$ achieves optimal performance in both IID and non-IID settings.
Remarkably, SatFed demonstrates performance enhancements over Ditto even without similarity edges.
This suggests that using the simply averaged peer models from satellite networks as the peer guide model $\Omega$ is still beneficial, although to a modest extent.
This enhancement is more significant in the non-IID scenario, reaching $1.17\%$, whereas negligible in the IID scenario.
This is likely because, in the IID setting, the global model excels in knowledge extraction, whereas its efficacy diminishes in the non-IID scenario, necessitating supplementary knowledge from peers.
Comparison of SatFed's accuracy with and without $\mathcal{A}_{sim}$ shows that the similarity capture in model transmissions within satellite networks boosts performance by $2.34\%$ in IID and $1.92\%$ in non-IID scenarios.
By mitigating the impact of outdated models on similarity updates and effectively guiding $\Omega$, $\mathcal{A}_{sim}$ significantly enhance the performance of local updates.

\subsubsection{Connection Edges}

\begin{figure}[t]
\centering
\begin{subfigure}{0.24\textwidth}
\includegraphics[width=\textwidth]{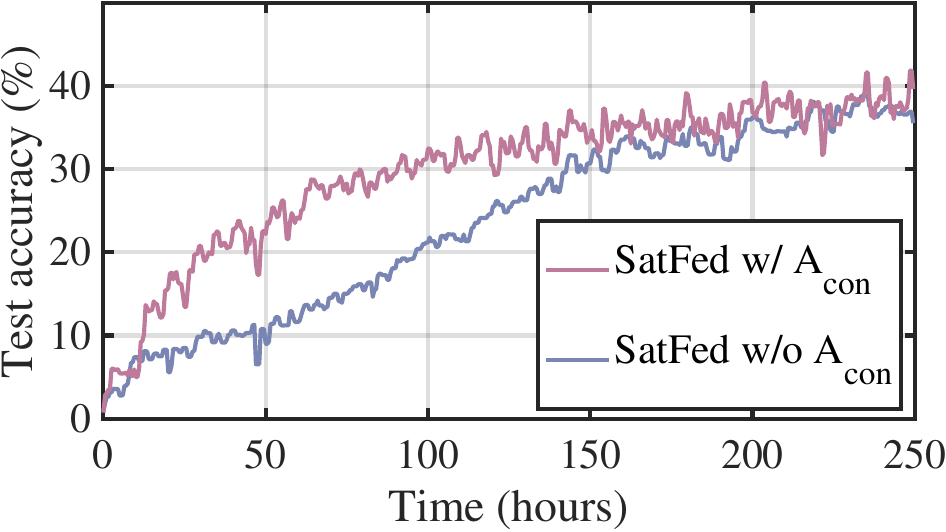}
    \caption{CIFAR-100 (IID).}
    \label{fig:ablation_con1}
\end{subfigure}
\hfill
\begin{subfigure}{0.24\textwidth}
\includegraphics[width=\textwidth]{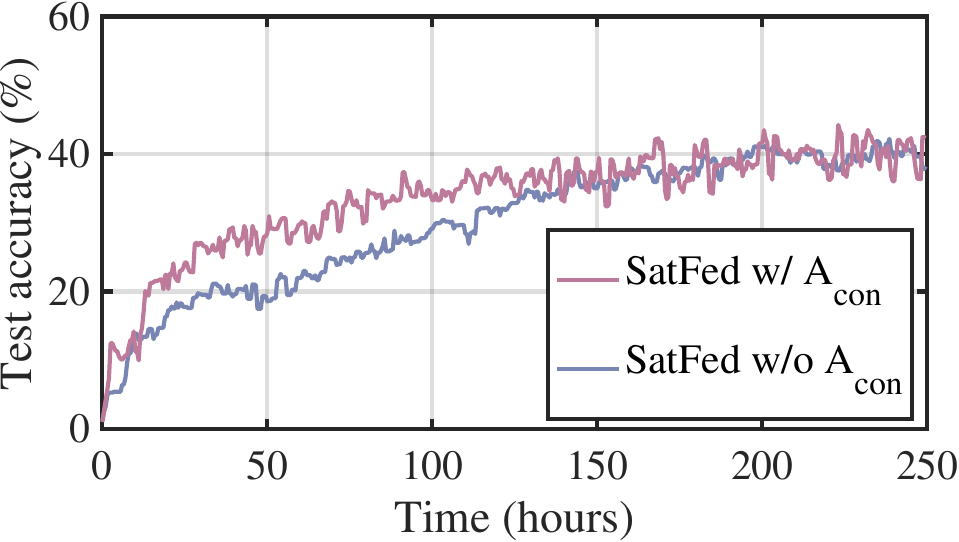}
    \caption{CIFAR-100 (non-IID).}
    \label{fig:ablation_con2}
\end{subfigure}

\caption{Ablation experiments for connection edges. To highlight the terrestrial bandwidth heterogeneity, all devices train with ideal computing power.}
\label{fig:ablation_con}
\vspace{-3.5ex}
\end{figure}

Fig.~\ref{fig:ablation_con} presents the impact of connection edges on CIFAR-100 dataset training performance.
Comparing with and without $\mathcal{A}_{con}$, it's evident that the edge connection's impact is most significant in the early stages of training. At this point, all devices' parameter updates are relatively similar. Models with poor terrestrial bandwidth can learn directly from the faster-progressing models through $\mathcal{A}_{con}$, significantly accelerating overall convergence.
However, in the later stages, due to the impact of data heterogeneity, the update directions of different devices diverge significantly, reducing the effectiveness of $\mathcal{A}_{con}$.


\subsubsection{Computation Edges}

\begin{figure}[t]
\centering
\begin{subfigure}{0.24\textwidth}
\includegraphics[width=\textwidth]{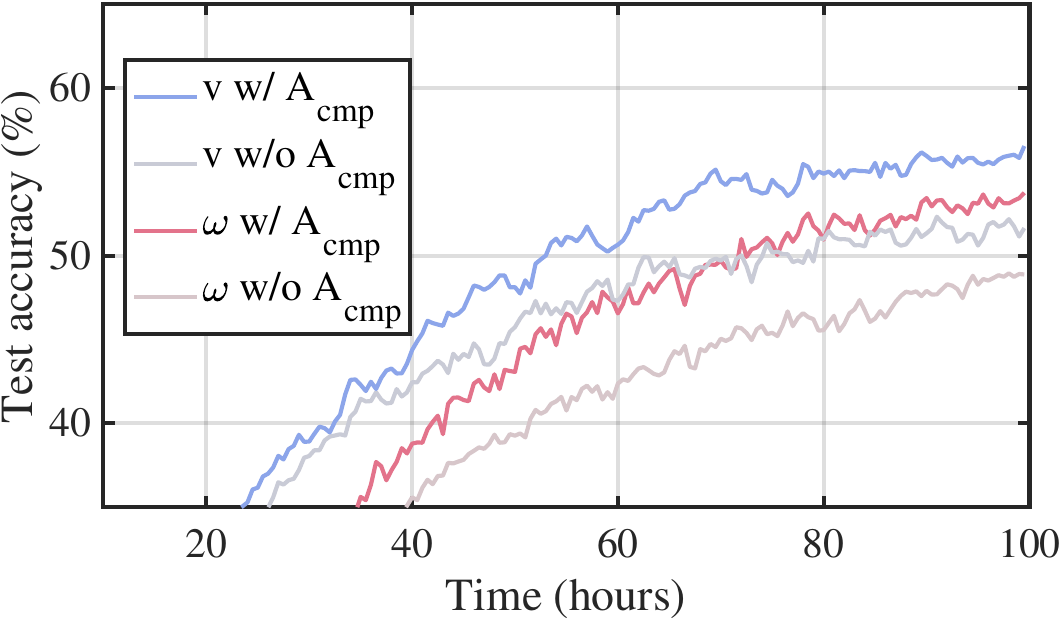}
    \caption{CIFAR-100 (IID).}
    \label{fig:ablation_comp1}
\end{subfigure}
\hfill
\begin{subfigure}{0.24\textwidth}
\includegraphics[width=\textwidth]{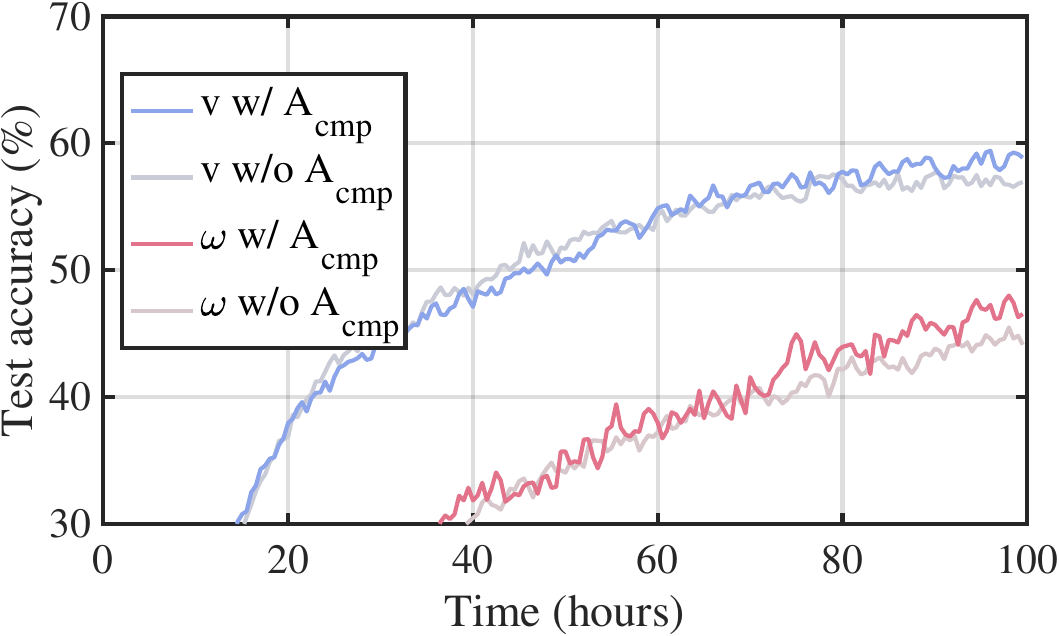}
    \caption{CIFAR-100 (non-IID).}
    \label{fig:ablation_comp2}
\end{subfigure}
\caption{Ablation experiments for computation edges. To emphasize the impact of computation heterogeneity, all operate within ideal terrestrial bandwidth.}
\label{fig:ablation_comp}
\vspace{-3.5ex}
\end{figure}

Fig.~\ref{fig:ablation_comp} illustrates the effect of computation edges on the SatFed training.
Clearly, $\mathcal{A}_{cmp}$ assisting in increasing the learning rate of undertrained models significantly enhances the performance of aggregated global models in IID scenarios. Although the improvement in non-IID scenarios is less pronounced, it still effectively enhances the injection of global knowledge into personalized models.
In summary, computation edges effectively enhance SatFed's performance across both IID and non-IID settings.
This underscores their ability to capture differences in local training progress and effectively address these differences through learning rate adjustments.

\section{Related Work}
\label{sec:relatedwork}

The system heterogeneity in FL, encompassing \textit{data}, \textit{bandwidth}, and \textit{computation} heterogeneity, is a well-established field with various approaches.
To address data heterogeneity challenges, Li \textit{et al.}~\cite{ditto} introduce personalized devices models, enhancing the fairness and robustness of the system.
Smith \textit{et al.}~\cite{NIPS2017_6211080f} adapts multitask learning for personalization by treating different devices as distinct learning tasks.
Ye \textit{et al.}~\cite{ye2023personalized} customizes the aggregation weights for each device by measuring the similarity between personalized models.
To tackle the issue of bandwidth heterogeneity, Nguyen \textit{et al.}~\cite{fedbuff} maintain a buffer for aggregation of device updates, reducing the performance degradation caused by outdated models.
Wang \textit{et al.}~\cite{wang2021resource} design a cluster-based FL mechanism with hierarchical aggregation, balancing communication resources by combining synchronous and asynchronous aggregations.
As for the uneven training issues caused by computational heterogeneity, Diao \textit{et al.}~\cite{diao2020heterofl} deploy device-specific local models adapted to match local computational resources and designs their aggregation method accordingly.
However, existing methods that address these FL system heterogeneities mainly focus on terrestrial PS communication architectures, overlooking the potential offered by emerging hybrid communication pathways such as satellite-assisted terrestrial networks.
This paper proposes novel solutions utilizing LEO satellite-assisted communication to tackle heterogeneity challenges, presenting an orthogonal approach compared to the aforementioned methods.

Satellite-terrestrial networks are an emerging and promising hybrid architecture with significant potential to alleviate terrestrial network transmission bottlenecks and improve overall transmission service quality~\cite{sun2022integrated}.
Cui \textit{et al.} utilize satellite networks to assist terrestrial IoT networks in reducing transmission latency~\cite{9955992}.
Ding \textit{et al.} design dynamic transmission and computation resource optimization for dense LEO satellite-assisted mobile-edge computing~\cite{ding2023dynamic}.
Although LEO satellites show potential to assist terrestrial communications, the exploration of FL in satellite-terrestrial hybrid network architectures remains largely uncharted.
The problem is non-trivial, requiring systematic design of model training, aggregation, and transmission for hybrid network architectures, while addressing satellite network topology, bandwidth issues, and various heterogeneity challenges in FL training.

\section{Conclusion}
\label{sec:conclu}

In this paper, we propose SatFed, a resource-efficient satellite-assisted FL framework designed to enhance training performance and convergence speed in complex heterogeneous environments.
Initially, SatFed designs a priority transmission queue based on model freshness differences, enabling devices in satellite networks with highly constrained transmission rates and complex topologies to receive near real-time peer models.
Furthermore, SatFed utilizes a multigraph to perceive in real-time the heterogeneity relationships among devices implied by models transmitted in satellite networks.
Finally, SatFed utilizes perceived heterogeneity relationships to help devices in constructing a peer guide model from satellite transmission, directing local updates.
Extensive experiments show that the SatFed framework outperforms state-of-the-art benchmarks.
In the future, we can explore more advanced distributed learning paradigms such as split learning~\cite{lin2024adaptsfl,lyu2023optimal}, and more powerful large language model~\cite{fang2024automated,qiu2024ifvit,hu2024agentscodriver,lin2024splitlora} deployments in satellite networks.




\ifCLASSOPTIONcaptionsoff
  \newpage
\fi



%


\bibliographystyle{IEEEtran}
\bibliography{NEWmybib}

\begin{thebibliography}{10}
\providecommand{\url}[1]{#1}
\csname url@samestyle\endcsname
\providecommand{\newblock}{\relax}
\providecommand{\bibinfo}[2]{#2}
\providecommand{\BIBentrySTDinterwordspacing}{\spaceskip=0pt\relax}
\providecommand{\BIBentryALTinterwordstretchfactor}{4}
\providecommand{\BIBentryALTinterwordspacing}{\spaceskip=\fontdimen2\font plus
\BIBentryALTinterwordstretchfactor\fontdimen3\font minus \fontdimen4\font\relax}
\providecommand{\BIBforeignlanguage}[2]{{%
\expandafter\ifx\csname l@#1\endcsname\relax
\typeout{** WARNING: IEEEtran.bst: No hyphenation pattern has been}%
\typeout{** loaded for the language `#1'. Using the pattern for}%
\typeout{** the default language instead.}%
\else
\language=\csname l@#1\endcsname
\fi
#2}}
\providecommand{\BIBdecl}{\relax}
\BIBdecl

\bibitem{wu2024s}
C.~Wu, H.~Cao, G.~Xu, C.~Zhou, J.~Sun, R.~Yan, Y.~Liu, and H.~Jiang, ``It's all in the touch: Authenticating users with host gestures on multi-touch screen devices,'' \emph{IEEE Trans. Mob. Comput.}, 2024.

\bibitem{10175391}
J.~Huang, D.~Li, C.~Huang, X.~Qin, and W.~Zhang, ``Joint task and data-oriented semantic communications: A deep separate source-channel coding scheme,'' \emph{IEEE Internet Things J.}, vol.~11, no.~2, pp. 2255--2272, 2024.

\bibitem{wu2022echohand}
C.~Wu, J.~Chen, K.~He, Z.~Zhao, R.~Du, and C.~Zhang, ``Echohand: High accuracy and presentation attack resistant hand authentication on commodity mobile devices,'' in \emph{Proc. CCS}, 2022, pp. 2931--2945.

\bibitem{lin2024efficient}
Z.~Lin, G.~Zhu, Y.~Deng, X.~Chen, Y.~Gao, K.~Huang, and Y.~Fang, ``Efficient parallel split learning over resource-constrained wireless edge networks,'' \emph{IEEE Trans. Mob. Comput.}, 2024.

\bibitem{liu2024sesame}
J.~Liu, J.~Ren, Y.~Zhang, S.~Yue, and Y.~Zhang, ``{SESAME: A Resource Expansion and Sharing Scheme for Multiple Edge Services Providers},'' \emph{{IEEE/ACM} Trans. Netw.}, pp. 1--16, 2024.

\bibitem{ren2022efficient}
J.~Ren, J.~Liu, Y.~Zhang, Z.~Li, F.~Lyu, Z.~Wang, and Y.~Zhang, ``An efficient two-layer task offloading scheme for mec system with multiple services providers,'' in \emph{Proc. of the 41st IEEE INFOCOM}, 2022, pp. 1519--1528.

\bibitem{lin2021spatial}
Z.~Lin, L.~Wang, B.~Tan, and X.~Li, ``Spatial-spectral terahertz networks,'' \emph{IEEE Trans. Wirel. Commun.}, vol.~21, no.~6, pp. 3881--3892, 2021.

\bibitem{wucong2024tifs}
C.~Wu, J.~Chen, Q.~Fang, K.~He, Z.~Zhao, H.~Ren, G.~Xu, Y.~Liu, and Y.~Xiang, ``Rethinking membership inference attacks against transfer learning,'' \emph{IEEE Trans. Inf. Forensics Secur.}, 2024.

\bibitem{huang2024d}
J.~Huang, K.~Yuan, C.~Huang, and K.~Huang, ``D$^{2}$-jscc: Digital deep joint source-channel coding for semantic communications,'' \emph{arXiv preprint arXiv:2403.07338}, 2024.

\bibitem{fang2024pacp}
Z.~Fang, S.~Hu, H.~An, Y.~Zhang, J.~Wang, H.~Cao, X.~Chen, and Y.~Fang, ``Pacp: Priority-aware collaborative perception for connected and autonomous vehicles,'' \emph{IEEE Trans. Mob. Comput.}, 2024.

\bibitem{hu2023towards}
S.~Hu, Z.~Fang, X.~Chen, Y.~Fang, and S.~Kwong, ``Towards full-scene domain generalization in multi-agent collaborative bird's eye view segmentation for connected and autonomous driving,'' \emph{IEEE Trans. Intell. Transp. Syst.}, 2024.

\bibitem{lin2022channel}
Z.~Lin, L.~Wang, J.~Ding, B.~Tan, and S.~Jin, ``Channel power gain estimation for terahertz vehicle-to-infrastructure networks,'' \emph{IEEE Commun. Lett.}, vol.~27, no.~1, pp. 155--159, 2022.

\bibitem{hu2024collaborative}
S.~Hu, Z.~Fang, Y.~Deng, X.~Chen, and Y.~Fang, ``Collaborative perception for connected and autonomous driving: Challenges, possible solutions and opportunities,'' \emph{arXiv preprint arXiv:2401.01544}, 2024.

\bibitem{dataprivacy}
L.~Marelli and G.~Testa, ``{Scrutinizing the EU General Data Protection Regulation},'' \emph{Science}, vol. 360, no. 6388, pp. 496--498, 2018.

\bibitem{li2024privacy}
Q.~Li, C.~Wu, J.~Chen, Z.~Zhang, K.~He, R.~Du, X.~Wang, Q.~Zhao, and Y.~Liu, ``Privacy-preserving universal adversarial defense for black-box models,'' \emph{arXiv preprint arXiv:2408.10647}, 2024.

\bibitem{rodio2023federated}
A.~Rodio, F.~Faticanti, O.~Marfoq, G.~Neglia, and E.~Leonardi, ``{Federated Learning under Heterogeneous and Correlated Client Availability},'' in \emph{Proc. of the 42nd IEEE INFOCOM}, 2023, pp. 1--10.

\bibitem{FedAvg}
B.~McMahan, E.~Moore, D.~Ramage, S.~Hampson, and B.~A. y~Arcas, ``{Communication-efficient Learning of Deep Networks from Decentralized Data},'' in \emph{Proc. of the 20th AISTATS}, 2017, pp. 1273--1282.

\bibitem{zhang2024fedac}
Y.~Zhang, H.~Chen, Z.~Lin, Z.~Chen, and J.~Zhao, ``Fedac: A adaptive clustered federated learning framework for heterogeneous data,'' \emph{arXiv preprint arXiv:2403.16460}, 2024.

\bibitem{lin2023fedsn}
Z.~Lin, Z.~Chen, Z.~Fang, X.~Chen, X.~Wang, and Y.~Gao, ``{FedSN: A General Federated Learning Framework over LEO Satellite Networks},'' \emph{arXiv preprint arXiv:2311.01483}, Nov. 2023.

\bibitem{chen2024gradient}
H.~Chen, Y.~Zhang, J.~Zhao, X.~Wang, and Y.~Xu, ``Gradient free personalized federated learning,'' in \emph{Proc. ICPP}, 2024, pp. 971--980.

\bibitem{wu2024wafbooster}
C.~Wu, J.~Chen, S.~Zhu, W.~Feng, K.~He, R.~Du, and Y.~Xiang, ``Wafbooster: Automatic boosting of waf security against mutated malicious payloads,'' \emph{IEEE Trans. Dependable Secure Comput.}, 2024.

\bibitem{sacco2021owl}
A.~Sacco, M.~Flocco, F.~Esposito, and G.~Marchetto, ``{Owl: Congestion Control with Partially Invisible Networks via Reinforcement Learning},'' in \emph{Proc. of the 40th IEEE INFOCOM}, 2021, pp. 1--10.

\bibitem{terrigrow}
P.~{\"O}hl{\'e}n, B.~Skubic, A.~Rostami, M.~Fiorani, P.~Monti, Z.~Ghebretensa{\'e}, J.~M{\aa}rtensson, K.~Wang, and L.~Wosinska, ``{Data Plane and Control Architectures for 5G Transport Networks},'' \emph{J. Light. Technol.}, vol.~34, no.~6, pp. 1501--1508, Feb. 2016.

\bibitem{lin2024split}
Z.~Lin, G.~Qu, X.~Chen, and K.~Huang, ``Split learning in 6g edge networks,'' \emph{IEEE Wirel. Commun.}, 2024.

\bibitem{yang2023detfed}
D.~Yang, W.~Zhang, Q.~Ye, C.~Zhang, N.~Zhang, C.~Huang, H.~Zhang, and X.~Shen, ``{DetFed: Dynamic Resource Scheduling for Deterministic Federated Learning over Time-Sensitive Networks},'' \emph{{IEEE} Trans. Mobile Comput.}, vol.~23, no.~5, pp. 5162--5178, 2024.

\bibitem{10040542}
J.~Xu, M.~A. Kishk, and M.-S. Alouini, ``{Space-Air-Ground-Sea Integrated Networks: Modeling and Coverage Analysis},'' \emph{{IEEE} Trans. Wireless Commun.}, vol.~22, no.~9, pp. 6298--6313, 2023.

\bibitem{zhai2023fedleo}
Z.~Zhai, Q.~Wu, S.~Yu, R.~Li, F.~Zhang, and X.~Chen, ``{{FedLEO}: An Offloading-assisted Decentralized Federated Learning Framework for Low Earth Orbit Satellite Networks},'' \emph{{IEEE} Trans. Mobile Comput.}, vol.~23, no.~5, pp. 5260--5279, 2024.

\bibitem{yuan2024satsense}
H.~Yuan, Z.~Chen, Z.~Lin, J.~Peng, Z.~Fang, Y.~Zhong, Z.~Song, and Y.~Gao, ``Satsense: Multi-satellite collaborative framework for spectrum sensing,'' \emph{arXiv preprint arXiv:2405.15542}, 2024.

\bibitem{zhang2022enabling}
Y.~Zhang, Q.~Wu, Z.~Lai, and H.~Li, ``{Enabling Low-Latency-Capable Satellite-Ground Topology for Emerging Leo Satellite Networks},'' in \emph{Proc. of the 41st IEEE INFOCOM}, 2022, pp. 1329--1338.

\bibitem{zhao2024leo}
Z.~Zhao, Z.~Chen, Z.~Lin, W.~Zhu, K.~Qiu, C.~You, and Y.~Gao, ``Leo satellite networks assisted geo-distributed data processing,'' \emph{arXiv preprint arXiv:2406.10856}, 2024.

\bibitem{yuan2023graph}
H.~Yuan, Z.~Chen, Z.~Lin, J.~Peng, Z.~Fang, Y.~Zhong, Z.~Song, X.~Wang, and Y.~Gao, ``Graph learning for multi-satellite based spectrum sensing,'' in \emph{Proc. ICCT}, 2023, pp. 1112--1116.

\bibitem{prevelance_of_satellite2}
T.~Ahmmed, A.~Alidadi, Z.~Zhang, A.~U. Chaudhry, and H.~Yanikomeroglu, ``{The Digital Divide in Canada and the Role of LEO Satellites in Bridging the Gap},'' \emph{IEEE Commun. Mag.}, vol.~60, no.~6, pp. 24--30, 2022.

\bibitem{10229104}
Z.~Lai, H.~Li, Y.~Wang, Q.~Wu, Y.~Deng, J.~Liu, Y.~Li, and J.~Wu, ``{Achieving Resilient and Performance-Guaranteed Routing in Space-Terrestrial Integrated Networks},'' in \emph{Proc. of the 42nd IEEE INFOCOM}, 2023, pp. 1--10.

\bibitem{CIFAR10}
A.~Krizhevsky, G.~Hinton \emph{et~al.}, ``Learning multiple layers of features from tiny images,'' 2009.

\bibitem{chen2022towards}
S.~Chen and B.~Li, ``{Towards Optimal Multi-Modal Federated Learning on Non-IID Data with Hierarchical Gradient Blending},'' in \emph{Proc. of the 41st IEEE INFOCOM}, 2022, pp. 1469--1478.

\bibitem{contacttime}
P.~Apollonio, C.~Caini, and M.~L{\"u}lf, ``{DTN LEO Satellite Communications Through Ground Stations and GEO Relays},'' in \emph{Proc. of the 5th PSATS 2013}, 2013, p. 1–12.

\bibitem{iperf}
A.~Tirumala, ``{Iperf: The TCP/UDP Bandwidth Measurement Tool},'' \emph{http://dast. nlanr. net/Projects/Iperf/}, 1999.

\bibitem{resnet}
K.~He, X.~Zhang, S.~Ren, and J.~Sun, ``{Deep Residual Learning for Image Recognition},'' in \emph{Proc. of the IEEE/CVF CVPR}, 2016, pp. 770--778.

\bibitem{yang2023efficient}
X.~Yang, J.~Chen, K.~He, H.~Bai, C.~Wu, and R.~Du, ``Efficient privacy-preserving inference outsourcing for convolutional neural networks,'' \emph{IEEE Trans. Inf. Forensics Secur.}, vol.~18, pp. 4815--4829, 2023.

\bibitem{heter1}
T.~Li, A.~K. Sahu, M.~Zaheer, M.~Sanjabi, A.~Talwalkar, and V.~Smith, ``{Federated Optimization in Heterogeneous Networks},'' \emph{Proc. of the 3rd MLSys}, pp. 429--450, 2020.

\bibitem{tan2022towards}
A.~Z. Tan, H.~Yu, L.~Cui, and Q.~Yang, ``{Towards Personalized Federated Learning},'' \emph{{IEEE} Trans. Neural Netw. Learn. Syst.}, vol.~34, no.~12, pp. 9587--9603, Mar. 2022.

\bibitem{liao2023adaptive}
Y.~Liao, Y.~Xu, H.~Xu, L.~Wang, and C.~Qian, ``{Adaptive Configuration for Heterogeneous Participants in Decentralized Federated Learning},'' in \emph{Proc. INFOCOM}, May. 2023.

\bibitem{basford2020performance}
P.~J. Basford, S.~J. Johnston, C.~S. Perkins, T.~Garnock-Jones, F.~P. Tso, D.~Pezaros, R.~D. Mullins, E.~Yoneki, J.~Singer, and S.~J. Cox, ``{Performance Analysis of Single Board Computer Clusters},'' \emph{Future Gener. Comput. Syst.}, vol. 102, pp. 278--291, 2020.

\bibitem{diao2020heterofl}
E.~Diao, J.~Ding, and V.~Tarokh, ``{Heterofl: Computation and Communication Efficient Federated Learning for Heterogeneous Clients},'' \emph{arXiv preprint arXiv:2010.01264}, Oct. 2020.

\bibitem{ditto}
T.~Li, S.~Hu, A.~Beirami, and V.~Smith, ``{Ditto: Fair and Robust Federated Learning through Personalization},'' in \emph{Proc. of the 38th ICML}, 2021, pp. 6357--6368.

\bibitem{pfedgraph}
R.~Ye, Z.~Ni, F.~Wu, S.~Chen, and Y.~Wang, ``{Personalized Federated Learning with Inferred Collaboration Graphs},'' in \emph{Proc. ICML}, Jul. 2023.

\bibitem{xiao2017fashion}
H.~Xiao, K.~Rasul, and R.~Vollgraf, ``Fashion-mnist: a novel image dataset for benchmarking machine learning algorithms,'' \emph{arXiv preprint arXiv:1708.07747}, 2017.

\bibitem{kingma2014adam}
D.~P. Kingma and J.~Ba, ``{Adam: A Method for Stochastic Optimization},'' \emph{arXiv preprint arXiv:1412.6980}, Dec. 2014.

\bibitem{asynchronousFL}
C.~Xie, S.~Koyejo, and I.~Gupta, ``Asynchronous federated optimization,'' \emph{arXiv preprint arXiv:1903.03934}, 2019.

\bibitem{NIPS2017_6211080f}
V.~Smith, C.-K. Chiang, M.~Sanjabi, and A.~S. Talwalkar, ``{Federated Multi-Task Learning},'' in \emph{Proc. of the 31th NIPS}, 2017, pp. 4427--4437.

\bibitem{ye2023personalized}
R.~Ye, Z.~Ni, F.~Wu, S.~Chen, and Y.~Wang, ``{Personalized Federated Learning with Inferred Collaboration Graphs},'' in \emph{Proc. ICML}, Jul. 2023.

\bibitem{fedbuff}
J.~Nguyen, K.~Malik, H.~Zhan, A.~Yousefpour, M.~Rabbat, M.~Malek, and D.~Huba, ``Federated learning with buffered asynchronous aggregation,'' in \emph{International Conference on Artificial Intelligence and Statistics}.\hskip 1em plus 0.5em minus 0.4em\relax PMLR, 2022, pp. 3581--3607.

\bibitem{wang2021resource}
Z.~Wang, H.~Xu, J.~Liu, H.~Huang, C.~Qiao, and Y.~Zhao, ``{Resource-Efficient Federated Learning with Hierarchical Aggregation in Edge Computing},'' in \emph{Proc. of the 40th IEEE INFOCOM}, 2021, pp. 1--10.

\bibitem{sun2022integrated}
Y.~Sun, M.~Peng, S.~Zhang, G.~Lin, and P.~Zhang, ``{Integrated Satellite-Terrestrial Networks: Architectures, Key Techniques, and Experimental Progress},'' \emph{IEEE Network}, vol.~36, no.~6, pp. 191--198, Jul. 2022.

\bibitem{9955992}
G.~Cui, P.~Duan, L.~Xu, and W.~Wang, ``{Latency Optimization for Hybrid GEO–LEO Satellite-Assisted IoT Networks},'' \emph{IEEE Internet Things J.}, vol.~10, no.~7, pp. 6286--6297, Apr. 2023.

\bibitem{ding2023dynamic}
C.~Ding, J.-B. Wang, M.~Cheng, M.~Lin, and J.~Cheng, ``{Dynamic Transmission and Computation Resource Optimization for Dense LEO Satellite Assisted Mobile-Edge Computing},'' \emph{IEEE Trans. Commun.}, vol.~71, no.~5, pp. 3087--3102, May. 2023.

\bibitem{lin2024adaptsfl}
Z.~Lin, G.~Qu, W.~Wei, X.~Chen, and K.~K. Leung, ``Adaptsfl: Adaptive split federated learning in resource-constrained edge networks,'' \emph{arXiv preprint arXiv:2403.13101}, 2024.

\bibitem{lyu2023optimal}
S.~Lyu, Z.~Lin, G.~Qu, X.~Chen, X.~Huang, and P.~Li, ``Optimal resource allocation for u-shaped parallel split learning,'' in \emph{Proc. Globecom Wkshps}, 2023, pp. 197--202.

\bibitem{fang2024automated}
Z.~Fang, Z.~Lin, Z.~Chen, X.~Chen, Y.~Gao, and Y.~Fang, ``Automated federated pipeline for parameter-efficient fine-tuning of large language models,'' \emph{arXiv preprint arXiv:2404.06448}, 2024.

\bibitem{qiu2024ifvit}
Y.~Qiu, H.~Chen, X.~Dong, Z.~Lin, I.~Y. Liao, M.~Tistarelli, and Z.~Jin, ``Ifvit: Interpretable fixed-length representation for fingerprint matching via vision transformer,'' \emph{arXiv preprint arXiv:2404.08237}, 2024.

\bibitem{hu2024agentscodriver}
S.~Hu, Z.~Fang, Z.~Fang, X.~Chen, and Y.~Fang, ``Agentscodriver: Large language model empowered collaborative driving with lifelong learning,'' \emph{arXiv preprint arXiv:2404.06345}, 2024.

\bibitem{lin2024splitlora}
Z.~Lin, X.~Hu, Y.~Zhang, Z.~Chen, Z.~Fang, X.~Chen, A.~Li, P.~Vepakomma, and Y.~Gao, ``Splitlora: A split parameter-efficient fine-tuning framework for large language models,'' \emph{arXiv preprint arXiv:2407.00952}, 2024.

\end{thebibliography}

\end{document}